\documentclass[12pt]{article}

\usepackage{arxiv}
\usepackage{authblk}
\usepackage{bigfoot}
\usepackage{etoolbox}
\usepackage{hyperref}

\usepackage{natbib}
\setcitestyle{authoryear,round,semicolon}

\usepackage{siunitx}
\usepackage{placeins} 
\usepackage{subfig}
\usepackage{amssymb}
\usepackage{amsmath}
\usepackage{multicol}
\usepackage{wrapfig}
\usepackage{cancel}
\usepackage{multirow}
\usepackage{graphicx}
\usepackage[dvipsnames,table,xcdraw]{xcolor}

\newcommand{\tild}{~}
\renewcommand{\figurename}{Fig.\tild}
\newcommand{\figref}{\figurename\ref}

\newcommand{\equationame}{Eq.\tild}
\newcommand{\equref}{\equationame\eqref}

\newcommand{\sectionname}{Sec.\tild}
\newcommand{\secref}{\sectionname\ref}

\DeclareNewFootnote{AAffil}[arabic]
\DeclareNewFootnote{ANote}[fnsymbol]

\makeatletter
\patchcmd\maketitle{\def\@makefnmark{\rlap{\@textsuperscript{\normalfont\@thefnmark}}}}{}{}{}
\makeatother

\makeatletter
\def\thanksAAffil#1{
  \footnotemarkAAffil\protected@xdef\@thanks{\@thanks%
        \protect\footnotetextAAffil[\the \c@footnoteAAffil]{#1}}%
}
\def\thanksANote#1{%
  \footnotemarkANote%
  \protected@xdef\@thanks{\@thanks%
        \protect\footnotetextANote[\the \c@footnoteANote]{#1}}%
}

\makeatother

\graphicspath{{figures/}}

\title{Solving adhesive rough contact problems \\with Atomic Force Microscope data}

\newbox{\orcid}\sbox{\orcid}{\includegraphics[scale=0.06]{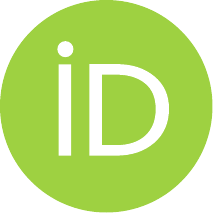}}

\author[1]{%
	\href{https://orcid.org/0000-0002-1528-4558}{\usebox{\orcid}\hspace{1mm}\textbf{Maria Rosaria Marulli}\thanks{\texttt{mariarosaria.marulli@imtlucca.it}, corresponding author}}%
}

\author[2,1]{%
	\href{https://orcid.org/0000-0001-8435-6466}{\usebox{\orcid}\hspace{1mm}\textbf{Jacopo Bonari}\thanks{\texttt{jacopo.bonari@dlr.de}}}%
}

\author[3]{%
	\href{https://orcid.org/0000-0003-1195-5630}{\usebox{\orcid}\hspace{1mm}\textbf{Pasqualantonio Pingue}\thanks{\texttt{pasqualantonio.pingue@sns.it}}}%
}

\author[1]{%
	\href{https://orcid.org/0000-0001-9409-9782}{\usebox{\orcid}\hspace{1mm}\textbf{Marco Paggi}\thanks{\texttt{marco.paggi@imtlucca.it}}}%
}

\affil[1]{IMT School for Advanced Studies Lucca, Piazza San Francesco 19, 56100 Lucca, Italy}

\affil[2]{Institute for the Protection of Terrestrial Infrastructures, German Aerospace Center (DLR), Rathausallee 12, 53757 Sankt Augustin, Germany}

\affil[3]{NEST - National Enterprise for nanoScience and nanoTechnology, Scuola Normale Superiore and Istituto Nanoscienze - CNR, Piazza San Silvestro 12, 56127 Pisa, Italy}

\hypersetup{
pdftitle={Solving adhesive rough contact problems with Atomic Force Microscope data},
pdfauthor={Maria Rosaria Marulli, Jacopo Bonari, Pasqualantonio Pingue, Marco Paggi},
pdfkeywords={First keyword, Second keyword, More keywords},
}

\begin{document}

\maketitle

\begin{abstract}
This study presents an advanced numerical framework that integrates experimentally acquired Atomic Force Microscope (AFM) data into high-fidelity simulations for adhesive rough contact problems, bridging the gap between experimental physics and computational mechanics. The proposed approach extends the eMbedded Profile for Joint Roughness (MPJR) interface finite element method to incorporate both surface topography and spatially varying adhesion properties, imported directly from AFM measurements. The adhesion behavior is modeled using a modified Lennard-Jones potential, which is locally parameterized based on the AFM-extracted adhesion peak force and energy dissipation data. The effectiveness of this method is demonstrated through 2D and 3D finite element simulations of a heterogeneous PS-LDPE (polystyrene matrix with low-density polyethylene inclusions) sample, where the bulk elastic properties are also experimentally characterized via AFM. The results highlight the significance of accounting for both surface adhesion variability and material bulk heterogeneity in accurately predicting contact responses.
\end{abstract}




\section{Introduction}

Contact problems between bodies are a fundamental topic in theoretical and applied mechanics and physics, with several critical applications for tribology in many engineering fields \citep{Vakis2018, Goryacheva2021}. Surface-surface interactions play a key role in stress transfer, friction, wear, heat, and electric conduction. Miniaturization observed nowadays in technology requires the development of high-fidelity contact mechanics simulations accounting for surface textures and microscopic roughness, which are observable over multiple length scales \citep{Paggi2020}. 

Adhesion plays a crucial role in micro-scale contact problems, where adhesive tractions influence contact mechanics to varying degrees \citep{Pastewka2014, Ciavarella2019}. In many cases, these tractions are relatively small compared to other contact contributions and can often be neglected. However, this is not always the case.
Adhesion becomes particularly important in micro- and nano-scale contacts, low-load conditions, and soft materials.
Thus, adhesive tractions are of great interest especially for the design of nano-devices, bio-systems, thin films, coatings, and layered structures, and affect the performance and durability of components in many engineering applications \citep{Komvopoulos2003, Ghatak2004, Spuskanyuk2008, Buskermolen2020,IPPOLITO2023}. 
Moreover, many real-world materials are heterogeneous, introducing further complexities in the contact response. Inhomogeneities can create stress concentrations, leading to fatigue crack nucleation and reduced fatigue strength. However, today, their impact on the adhesive contact of rough surfaces remains challenging to be predicted using numerical methods, often requiring assumptions on the surface roughness and/or the adhesive behavior \citep{KESARI2011,JOE2018, PUTIGNANO2012p, ZHAO2022, Sanner2024}.

In this context, numerical solution methods based on the Boundary Element Method (BEM) have been widely exploited \citep{ Andersson1983, Xu2019, Bemporad2020,PUTIGNANO2012, MARULLI2025} since they benefit from the sole discretization of the boundary and not of the bulk. However, this methodology is intrinsically limited to problems where the Green functions, relating the displacements of the points belonging to the boundary to the applied contact tractions, are available, and to linear problems where the principle of superposition can be invoked. Therefore, although some significant efforts have been made to extend and apply BEM to adhesive contact \citep{Medina2014, Popov2017, Rey2017,SANNER2022}, viscoelasticity \citep{Putignano2015, Menga2022}, the underlying model assumptions preclude its straightforward generalization to nonlinear interface and bulk constitutive relations, and also to multi-field coupled problems. 
Another possible solution strategy consists of semi-analytical solutions as in \citep{Chen2024}, where they analyzed adhesive contact behavior in heterogeneous elastic materials with engineered rough surfaces. The proposed equivalent inclusion method (EIM), however, assumes idealized inhomogeneities, which may not fully capture complex material structures with graded or anisotropic properties, and the rough surfaces are modeled based on the statistical description of the height field distributions. 

To overcome the above limitations, the Finite Element Method (FEM) would be the natural remedy. However, due to the higher computational cost associated with the discretization of the bulk and the complexity in meshing nonplanar interfaces, its application to contact problems with roughness has been primarily confined to a few instances \citep{Pei2005, Reinelt2009, Yastrebov2011, Radhakrishnan2020, AFFERRANTE2023, Luo2025}. Moreover, the convergence of contact search algorithms in multi-scale roughness can be problematic. 

To address the challenges outlined above, a finite element discretization technique relying on the eMbedded Profile for Joint Roughness (MPJR) interface finite element, and the associated contact solution scheme, have been firstly proposed by \cite{Paggi2018} and extended in \citep{Bonari2020, Bonari2021, Bonari2022} to frictional problems.
This novel framework has been developed to solve high-fidelity contact mechanics simulations accounting for surface textures and microscopic roughness. An interface characterized by any arbitrarily complex shape can be globally discretized as smooth. At the same time, any deviation from planarity is embedded into the MPJR interface finite elements to be used as a correction of the normal gap function computed if the surfaces were flat. It applies to both rigid-deformable and deformable-deformable solids in contact and is particularly efficient from the computational point of view for the following principal reasons: $(i)$ the interface is globally discretized as nominally flat; $(ii)$ any height field perturbing the planarity can be incorporated, either given by a continuous analytical function when available or directly provided by discrete profilometric data; $(iii)$ no smoothing or regularization of the height field is required; $(iv)$ contact search algorithms can be fully avoided by considering a fixed pairing of nodes at the interface. 

Moreover, the methodology can be extended to other nonlinear phenomena or multi-field problems. It has been applied to frictional contacts in partial slip \citep{Bonari2021} and full sliding regime \citep{Bonari2020} for viscoelastic solids. Moreover, the method has been extended to 3D, including friction and adhesion \citep{Bonari2022}, demonstrating its efficiency through specific benchmark tests on spherical indenters, wavy profiles, and fractal surfaces, comparing the results with theoretical solutions and BEM predictions. The solution of contact-induced fracture phenomena has been treated by \citep{Marulli2023}, where the MPJR method combined with the phase field approach to brittle fracture for the substrate has been used for the simulation of smooth and rough spherical indentation tests, with a very good agreement with available experimental trends. 

The present study extends and exploits the above  methodology to solve complex contact problems involving not only experimentally acquired surface topography, but also the material heterogeneity and the spatial variation of the adhesive field on the substrate surface. The MPJR framework is therefore extended to consider, in addition to the surface roughness, multiple field variables at the interface level, in particular, collected with an Atomic Force Microscope (AFM).

\begin{figure}[h]
    \centering
    \includegraphics[width=0.4\textwidth]{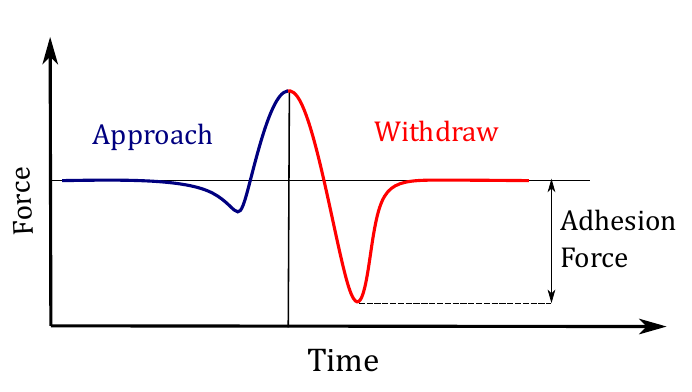}\qquad
    \includegraphics[width=0.4\textwidth]{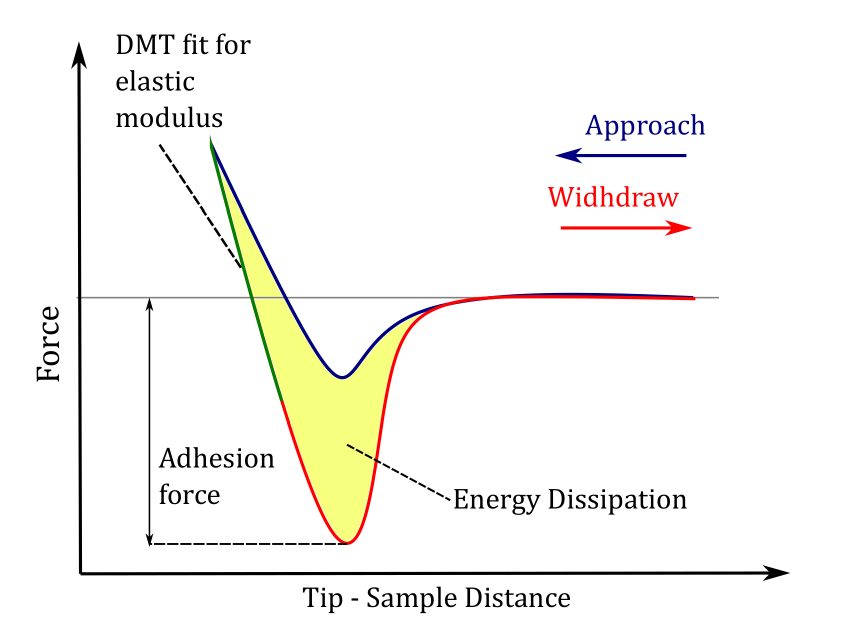}
    \caption{AFM schemes: (a) force-time plot of the scanning; (b) force-distance curve with main variables. }
    \label{fig:afm}
\end{figure}

AFMs are primarily used to obtain topographic information on solid surfaces at the nanometer scale and below. A sharp tip mounted on a cantilever touches the surface, which causes the cantilever to deflect. This deflection is proportional to the forces acting between the microscope tip and the investigated surface \citep{Haba2014}. During the surface scanning, the real-time force-displacement curve is recorded for each surface point (see \figref{fig:afm}), and surface properties such as morphology, adhesive force, elastic modulus, surface deformation, can be derived from the force-displacement curve at the same time. Besides the sample height $Z$, the most important measured for this work are the elastic modulus, the adhesion force (which is the maximum force reached during the withdraw curve), and the energy dissipated during the approach-withdraw cycle, represented by the yellow area in \figref{fig:afm}b. The elastic modulus is determined by fitting the retract force-distance curves to a Derjaguin-Muller-Toporov (DMT) model \citep{Derjaguin1975}. This model approximates the AFM tip as a sphere pressing into a flat surface and links the contact force and the surface deformation to the material modulus. 

The present work describes how the punctual values of adhesion peak force and energy dissipation collected with the AFM can be fully embedded in the MPJR interface finite elements, paving the way to a full integration of AFM data into finite element analysis software for high-fidelity contact mechanics predictions, bridging the gap between materials science and physics experiments and numerical simulation. The numerical framework is presented in \secref{sec:numerical} and applied to a 3D contact problem in \secref{sec:3dcontact}. The AFM also provides information related to the elastic modulus of the analyzed material. This feature is not directly included in the MPJR finite element since it is a bulk property, but the effect of the material heterogeneity on the contact solution is discussed in \secref{sec:elastic_mod}, proving the ability of the proposed numerical method for the analysis of the heterogeneous material.

\section{Numerical framework} \label{sec:numerical}

The derivation of the MPJR interface finite elements is briefly summarized in this section (the interested reader can refer to \citep{Paggi2018, Bonari2022} for more details), focusing on how not only the experimental height field but also the adhesion and dissipation fields have been exploited to formulate the interface constitutive model, allowing for an accurate solution of the contact problem. 

\begin{figure}[h]
    \centering
    \includegraphics[width=0.75\linewidth]{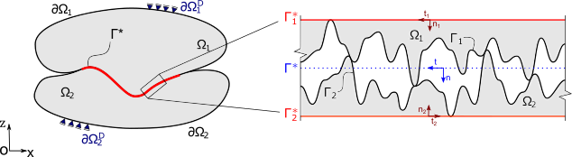}
    \caption{Solid domains $\Omega_{i=1,2}$ coming into contact at the nominally smooth interface $\Gamma^*$.}
    \label{fig:domains}
\end{figure}

Let the contact problem be described by two domains $\Omega_{i=1,2}$ separated by an interface $\Gamma^*$ with opposite boundaries $\Gamma_{i=1,2}^*$ nominally smooth but microscopically embedding the rough surfaces $\Gamma_{i=1,2}$, as represented in \figref{fig:domains} and by the kinematic quantities governing the contact problem. 

\begin{figure}
    \centering
    \includegraphics[width=0.75\linewidth]{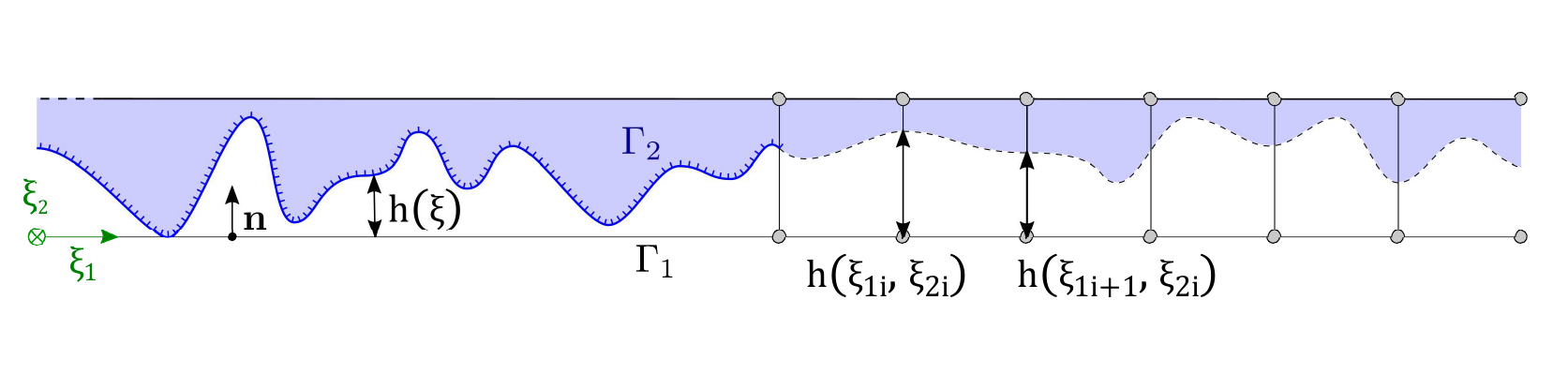}
    \caption{Nominally smooth interface $\Gamma^*$ with embedded rough surface defined by the roughness function $h(\boldsymbol{\xi})$, and its discretization in MPJR interface finite elements. }
    \label{fig:interface}
\end{figure}

In general, two indenting rough surfaces can be modeled in the proposed numerical framework by considering their \textit{composite topography}, see \citep{Paggi2018}. In this work, for simplicity, the case of a smooth surface $\Gamma_1$ in contact with a rough surface $\Gamma_2$ is considered. The geometrical difference between $\Gamma_2$ and $\Gamma_2^*$ is mathematically described by a function $h(\boldsymbol{\xi})$ where $\boldsymbol{\xi} = (\xi_1,\xi_2)^T$ is a curvilinear coordinate defining a point-wise correspondence with the coordinates of the same point in the global reference system. The surface elevation is computed with respect to a reference smooth surface parallel to the average plane with datum set in correspondence of its deepest valley, see \figref{fig:interface}.

The contact interactions between the solids can be described by the relation between the gap vector $\boldsymbol{g}=\{g_n, g_{t1}, g_{t2}\}^T$ at the contact interface $\Gamma$ and the contact tractions $\boldsymbol{\tau}=\{p_n, \tau_1. \tau_2 \}^T$. For simplicity, a frictionless contact problem is considered here, focusing on the derivation of the normal components. The frictional contribution can be added to the framework as in \citep{Bonari2021}.

The contact interface contribution to the weak form of the equilibrium equation can be written as:
\begin{equation}\label{contribution}
\Pi_{\Gamma^*}=\int_{\Gamma^*} p_n (\mathbf{u}) \delta g_n (\mathbf{v}) \, \mathrm{d} \Gamma.
\end{equation}

\begin{figure}[h]
    \centering
    \includegraphics[width=0.7\linewidth]{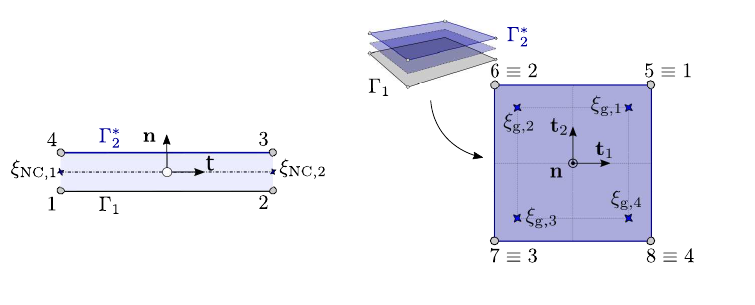}
    \caption{2D and 3D interface finite elements.}
    \label{fig:elements}
\end{figure}

The numerical solution of the contact problem in the finite element framework requires the discretization of the two bodies and of the contact interface into finite elements. In this work, the bulk has been discretized by linear isoparametric finite elements, even though there is no restriction on the finite element topology, provided that it is consistent with that of the MPJR interface finite element used, which are represented in \figref{fig:elements} for the 2D and 3D cases. 

Introducing a vector of nodal displacements $\mathbf{u} =
[u_1, v_1, . . . , u_8, v_8]^T$ (in 3D), the gap vector $\boldsymbol{g}$ can be computed as:
\begin{equation}
    \boldsymbol{g} = \mathbf{Q} \mathbf{N}\mathbf{L}\mathbf{u},
\end{equation}
where $\mathbf{L}$ is a linear operator for computing the relative displacements across the interface, $\mathbf{N}$ is the shape functions matrix, and $\mathbf{Q}$ is a rotation matrix for transforming displacements from the global to the local reference frame of the interface finite element (see \citep{Paggi2018} for the operators' derivation).

The original deviation from planarity between the smoothed and the actual geometry of the contact surface can be restored by a suitable correction of the normal component of the gap: $g_n^*=g_n+h(\boldsymbol{\xi})$, where $h(\boldsymbol{\xi})$ establishes a one-to-one correspondence of the rough surface with the coordinates of the same point in the global reference system.

In the case of AFM data, the surface elevation field is defined as $z_{ij}$ at each scanning point $(x_i, x_j)$. This elevation field is introduced in the MPJR finite element considering the corresponding modified normal gap computed at each integration point as:
\begin{equation}\label{eq:gap}
   g_n^\ast (x_i, x_j) = g_n(x_i, x_j) + z_{ij}. 
\end{equation}

Using the modified gap in the derivation of the stiffness matrix of the system allows us to account for the complex geometry without considering it explicitly during the FE discretization process. 

When an adhesive contact problem is considered, the constitutive relation in the normal direction can be written using a Lennard-Jones potential-like relationship \citep{Sauer2009, Yu2004}:
\begin{equation}\label{eq:len-jo}
    p_n=\frac{A_N}{6 \pi g_0^3} \left[ \left(\frac{g_0}{g_n^*}\right)^9 - \left(\frac{g_0}{g_n^*}\right)^3 \right]
       = \frac{8\Delta\gamma}{3 g_0} \left[ \left(\frac{g_0}{g_n^*}\right)^3 - \left(\frac{g_0}{g_n^*}\right)^9 \right],
\end{equation}
where $A_N$ is the Hamaker constant; $g_0$ is the equilibrium spacing between the two half-spaces; and $\Delta \gamma$ is the adhesion energy per unit area. Note that here and in the remainder of the manuscript, positive values of $p_n$ are associated with attractive tractions (adhesion); in contrast, repulsive tractions have a negative sign, contrary to what is generally assumed for the AFM curve in \figurename\ref{fig:afm}.

The AFM provides the value of the dissipated energy during a tip approach-withdrawing cycle, given by the yellow area in \figref{fig:afm}. Under the hypothesis that the area under the approach curve is negligible, the dissipated energy can be considered equal to the adhesive energy $\Delta \gamma$.
The maximum adhesive traction, $p_{\max}$, provided by AFM can be exploited to calculate the value of $g_0$ in \equref{eq:len-jo} as follows. The stationary point $g_{\max}$ of the equation can be expressed in function of $g_0$ as:
\begin{equation}
    g_{\max} = \sqrt[6]{3}g_0.
\end{equation}
If this value is introduced in Eq.~\eqref{eq:len-jo}, then a functional relationship can be derived linking $p_{\max}$ and $g_0$:
\begin{equation}
   g_0 = \frac{16}{9\sqrt{3}}\frac{\Delta\gamma}{p_{\max}}.
\end{equation}
Finally, this value can be substituted into \equref{eq:len-jo} to derive a potential that is only function of $p_{\max}$ and $\Delta \gamma$, that is, the quantities acquired by AFM scan. The very same equation can be expressed as:
\begin{equation}\label{eq:len-john-power}
    p_n = - (a_1 g_n^{*b_1} - a_2 g_n^{*b_2}),
\end{equation}
where:
\begin{equation}
    a_1 = 3.284\frac{\Delta \gamma^9}{p_{\max}^8}, \qquad
    a_2 = 2.809\frac{\Delta \gamma^3}{p_{\max}^2}, \qquad b_1=-9, \qquad b_2=-3.
\end{equation}

\begin{figure}[h]
    \centering
    \includegraphics[width=0.7\textwidth]{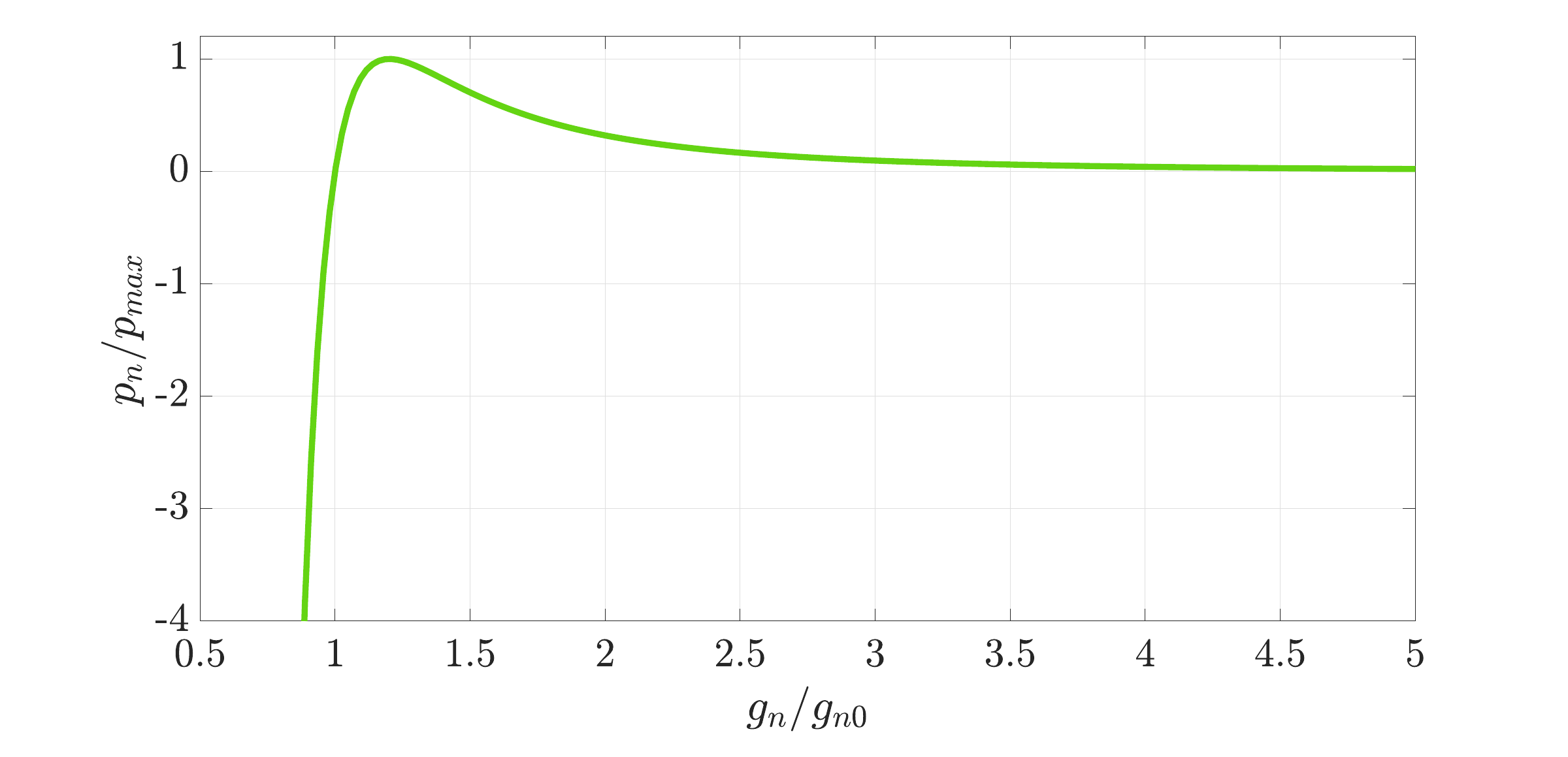}
    \caption{Lennard-Jones potential employed as the interface constitutive law.}
    \label{fig:len-jo}
\end{figure}

The resulting Lennard-Jones potential plotted in Fig.~\ref{fig:len-jo} is theoretically defined for any value of $g_n^*$, presenting an asymptote in $g_n^*=0$.
Simulating adhesive contact problems using this model presents several challenges, primarily due to the highly nonlinear nature of the Lennard–Jones potential. One of the main difficulties is the absence of an undeformed equilibrium position for two interacting deformable bodies. According to the model, as long as the bodies remain within a finite distance, interaction forces arise, leading to continuous deformation.
Another significant challenge is the appearance of attractive instability in cases of strong adhesion. This instability, commonly referred to as jump-to-contact and jump-off-contact, can be particularly problematic in quasi-static simulations. To mitigate this issue, arc-length methods or dynamic simulation techniques can be used \citep{Sauer2009}.
Additionally, the tangent matrix may become ill-conditioned as $g_n^* \to 0 $, because the Lennard-Jones potential and the corresponding contact stiffness tend to infinity. This issue is particularly pronounced when large load increments are used or when finite element discretization is too coarse, especially in the presence of strong adhesion, as the contact stiffness scales with $A_N$. To address this, smaller load increments and finer element discretization can be utilized. 

In the present numerical formulation, the relationship  in \equationame\eqref{eq:len-john-power} has been regularized by $(i)$ considering that the adhesive contribution can be neglected for high values of the gap and $(ii)$ removing the asymptote at $g_n^*=0$.
The first regularization approximates the Lennard-Jones curve with a linear law between $g_{nc1}$ and $g_{nc2}$. These two values have been set by computing analytically the area under the Lennard-Jones curve. The gap value $g_{nc1}$ is set such that the area in the range $[g_{n0}, g_{nc1}]$ is equal to $99\%$ of the total area. The value $g_{nc2}$ is computed by setting the area of the approximating linear branch equal to $1\%$ of the total area. Regarding the asymptote, a linear approximation is introduced: for $g_n^* \le g_{n0}$, the normal contact traction is computed using the analytical tangent to the Lennard-Jones potential at $g_n^*=0$ (noted as $k_0$). The tangent can be further adjusted to prevent ill-conditioning problems for $g_n^* \to 0$ by multiplying its value with a user-defined factor, denoted as $k_t$. In the numerical simulations presented in the remainder of the paper, this value is set equal to $100 E/L$ where $E$ is the elastic modulus, and $L$ is the characteristic size of the sample.

\begin{figure}
    \centering
    \includegraphics[width=0.7\linewidth]{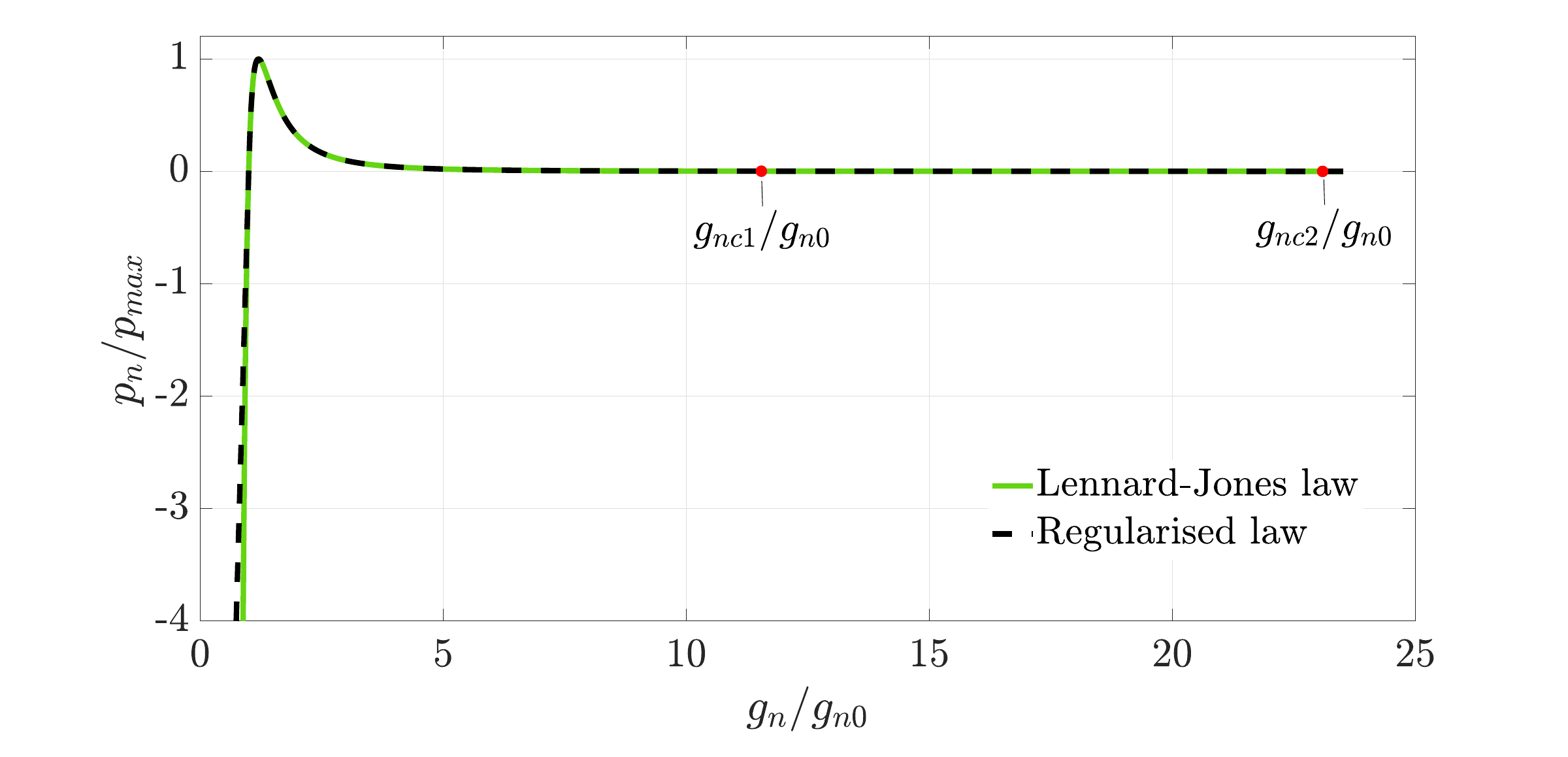}
    \caption{Regularization of the Lennard-Jones potential for the interface finite element.}
    \label{fig:int_law_appr}
\end{figure}

The regularized version of the Lennard-Jones (plotted in \figref{fig:int_law_appr} as the dashed black line) reads:
\begin{equation}\label{eq:interf_appr}
    p_n=\begin{cases}
       \vphantom{\frac{0}{0}}  k_t \, k_0 g_n^* \qquad & \text{if} \quad g_n^* \le g_{n0}, \\
        
       \vphantom{\frac{0}{0}} - (a_1 g_n^{*b_1} - a_2 g_n^{*b_2})  \qquad & \text{if} \quad g_{n0} < g_n^* \leq g_{nc1}, \\
        
        \dfrac{a_1 g_{nc1}^{b_1}-a_2 g_{nc1}^{b_2}}{g_{nc2}-g_{nc1}} (g_n^*-g_{nc2}) \qquad & \text{if} \quad g_{nc1}< g_n^* \leq g_{nc2}, \\
        
        \vphantom{\frac{0}{0}} 0 & \text{if} \quad g_n^* > g_{nc2}.
    \end{cases}
\end{equation}

Compared to the analysis by \cite{Bonari2020} of adhesive contact problems, the parameters in \eqref{eq:interf_appr} depend on the coordinate of the interface integration point.  
To minimize access to external files, the surface elevation field and the Lennard-Jones parameters are stored in the finite element routine based on the correspondence between the AFM data spatial coordinates and the coordinates of the interface finite element, using three history variables for each integration point; this operation is done only once at the initialization of the FE simulation. 

As in the previous authors' contributions, the interface finite element is coded as a user element in the finite element analysis program FEAP8.6 \citep{Zienkiewicz2013}. A full Newton-Raphson iterative-incremental scheme is employed to solve the nonlinear system of algebraic equations stemming from the FE discretized weak form.

\section{3D adhesive rough contact simulation}\label{sec:3dcontact}

The capabilities of the proposed numerical framework are tested in this section using actual experimental data provided by the NEST laboratory of Scuola Normale Superiore (Pisa, Italy) using their BRUKER "DIMENSION ICON" Atomic Force Microscopy (AFM) system, tipically employed for characterization of surfaces at the nanoscale. 

\begin{figure}[h]
    \centering
    \includegraphics[width=0.7\textwidth]{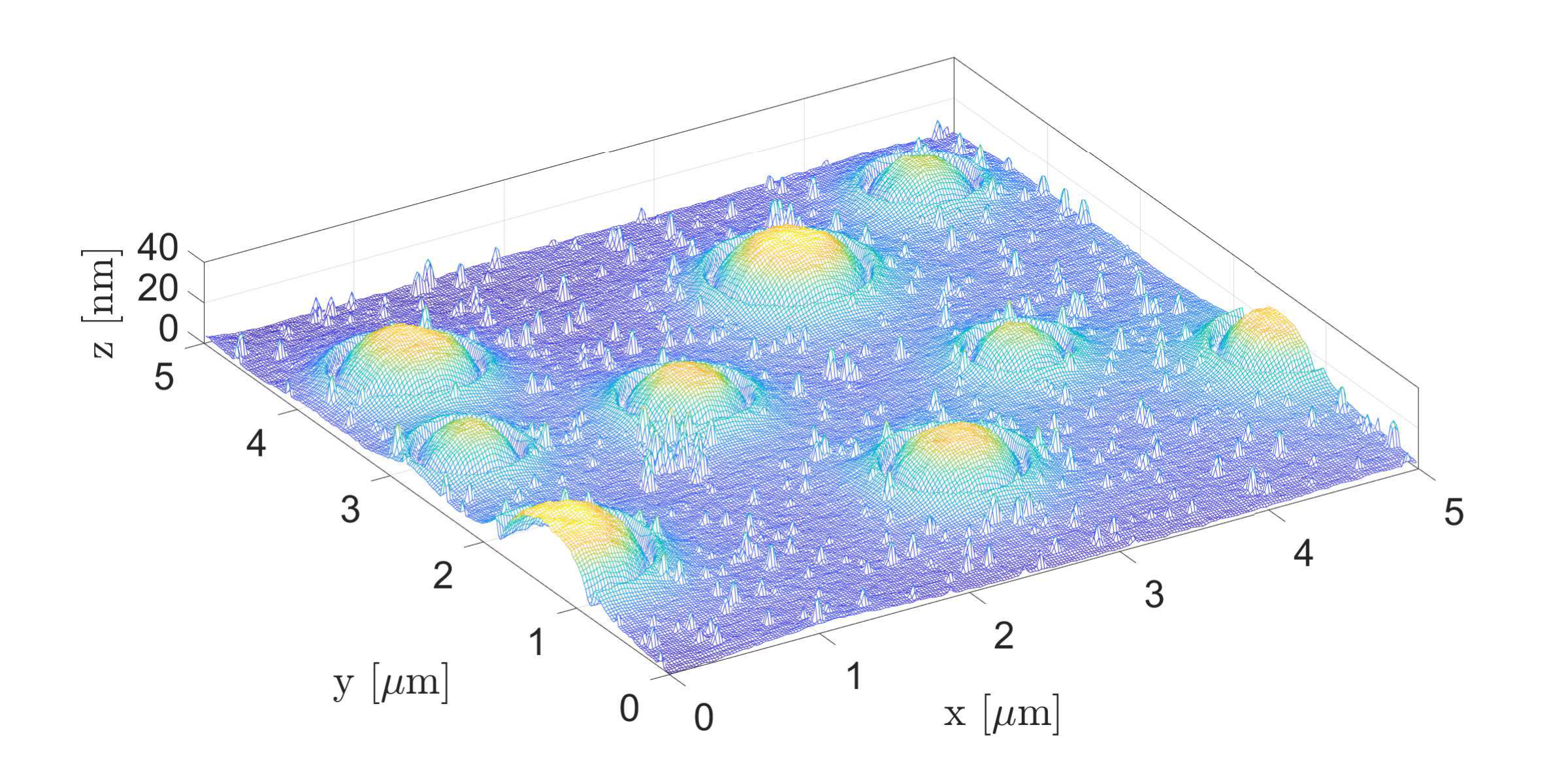}
    \caption{Height field of the PS-LDPE sample where the LDPE semi-spherical inclusions are clearly visible. }
    \label{fig:topo}
\end{figure}

The material used for the first example consists of a PS-LDPE sample composed of a polystyrene (PS) matrix and low-density polyethylene (LDPE) as a doping component. The morphology of the sample is shown in \figref{fig:topo} and represents a scanned surface of $ 5 \times 5\tild\si{\mu m^2}$ with 256 points per side. The maximum elevation is $h_\text{max}= 33 \tild\si{nm}$ and the root-mean-square of the heights field is $h_\text{rms} =24.8 \tild\si{nm}$. The topography has been used without any filtering. The presence of spikes and hard discontinuities on the surface represents a challenge for traditional numerical methods, allowing us to test the real capabilities of the MPJR framework.  

\begin{figure}[h!]
    \centering    \includegraphics[width=0.3\textwidth]{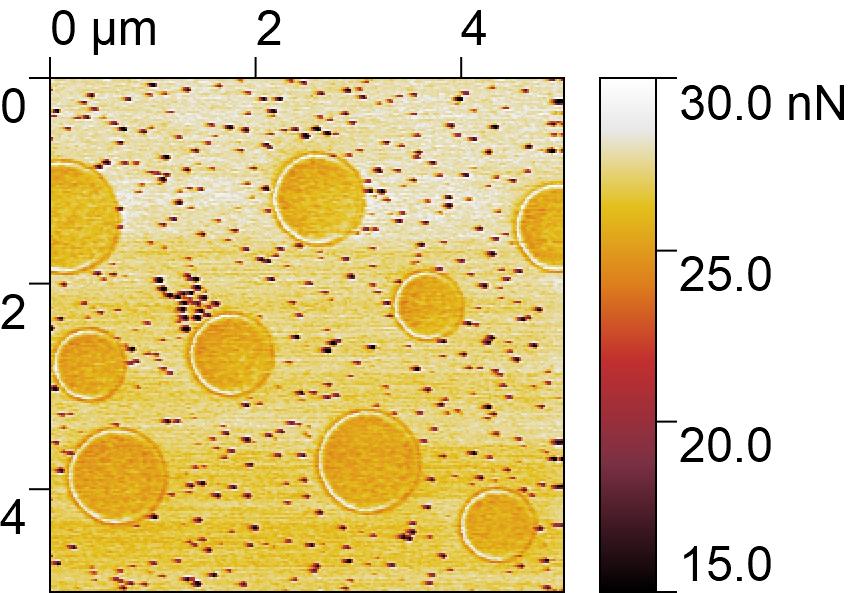} \qquad  \qquad  \includegraphics[width=0.3\textwidth]{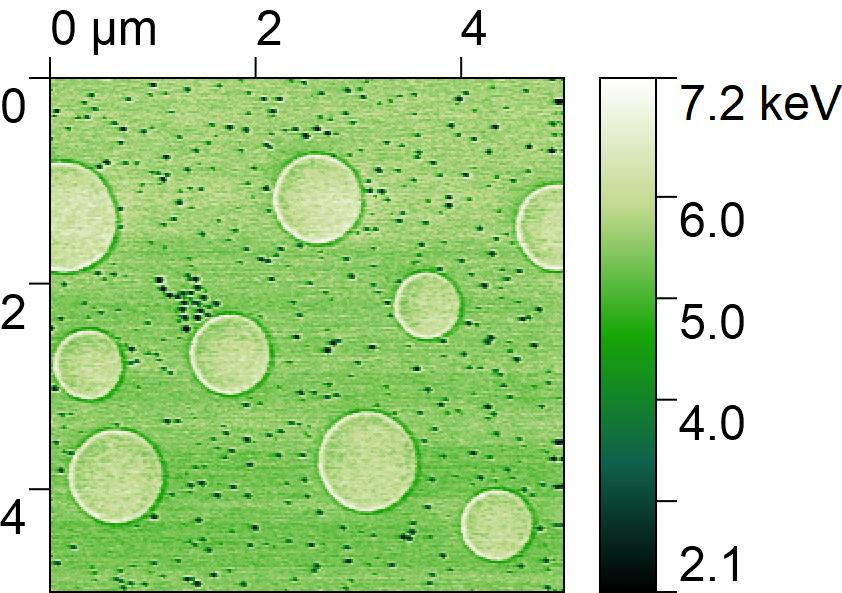}
    \caption{Adhesive force field on the left, and energy dissipation on the right, for the PS-LDPE sample.}
    \label{fig:adh}
\end{figure}

The variation of the adhesive peak force and energy dissipation on the sample surface is shown in \figref{fig:adh}.

\begin{figure}[h!]
    \centering    \includegraphics[width=0.25\textwidth]{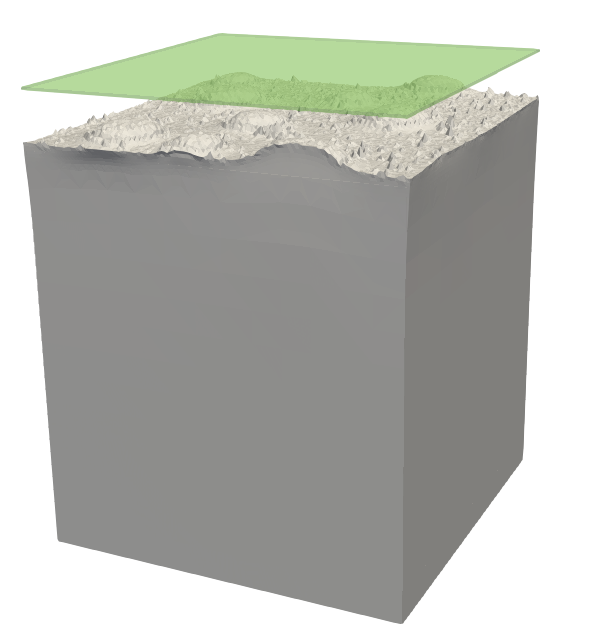} \qquad \qquad \qquad
    \includegraphics[width=0.25\textwidth]{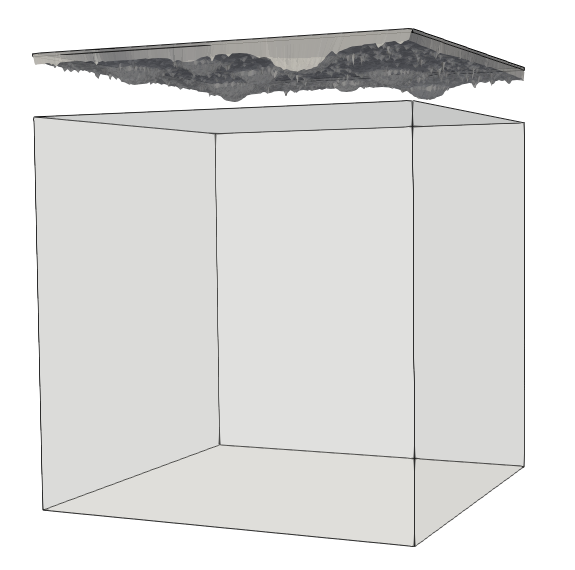}
    \caption{Benchmark test: contact problem between the PS-LDPE sample and a flat rigid indenter, on the left, and the equivalent problem on the right.}
    \label{fig:tests}
\end{figure}

As a benchmark problem to demonstrate the capabilities of the described numerical framework, we simulate the contact problem between the deformable PS-LDPE sample presented above with a rigid flat indenter, as shown in \figref{fig:tests}(a). Under the hypothesis of small displacements, this problem is equivalent to the case of a rigid rough indenter having the sample height field indenting a deformable substrate with the PS-LDPE mechanical properties, as demonstrated by \cite{Barber2018}. The elevation field of the rough indenter has been obtained by computing the composite topography as described in \cite{Paggi2018} and shown in \figref{fig:tests}(b).

\begin{figure}[h!]
    \centering    \includegraphics[width=0.5\textwidth]{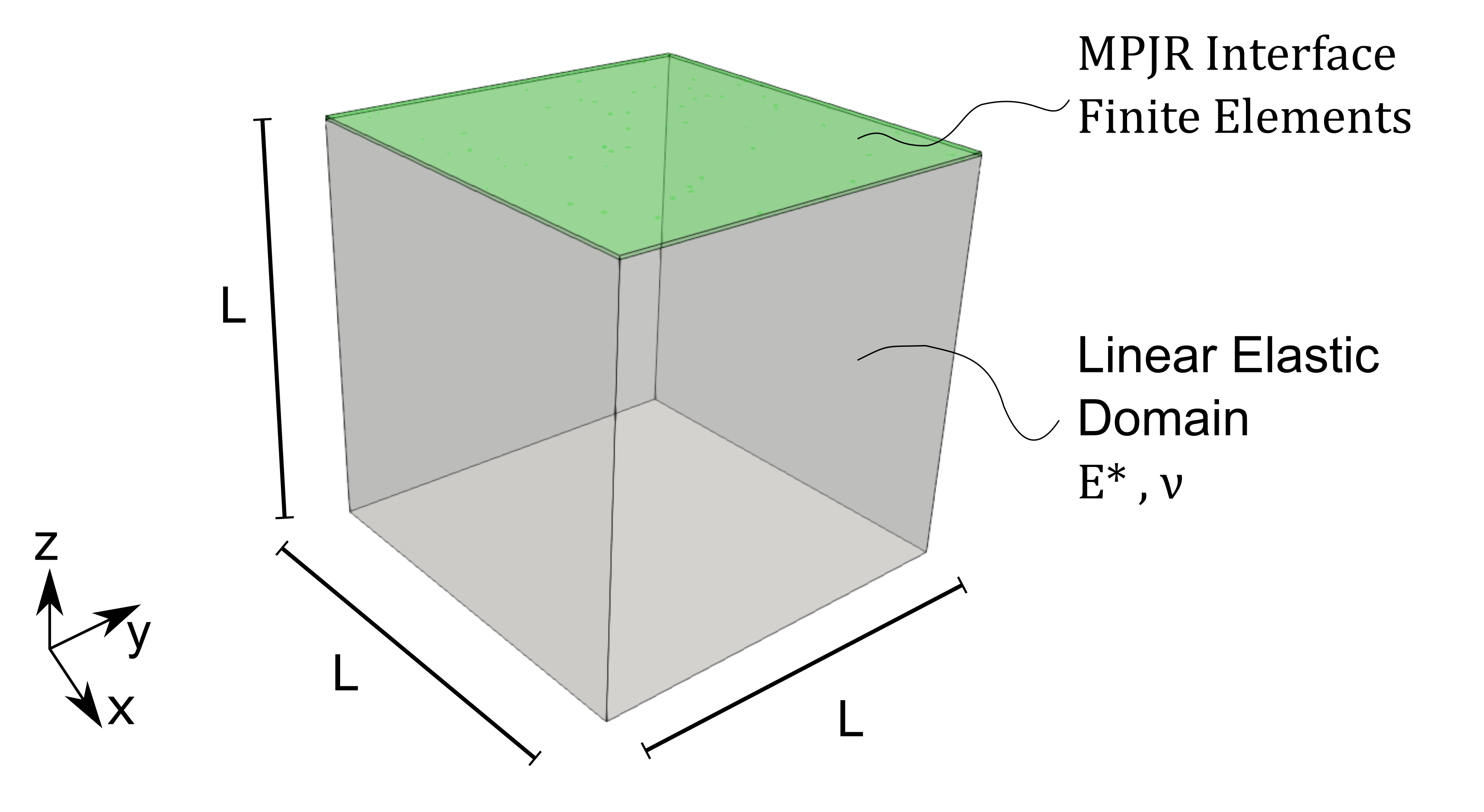}
    \includegraphics[width=0.3\textwidth]{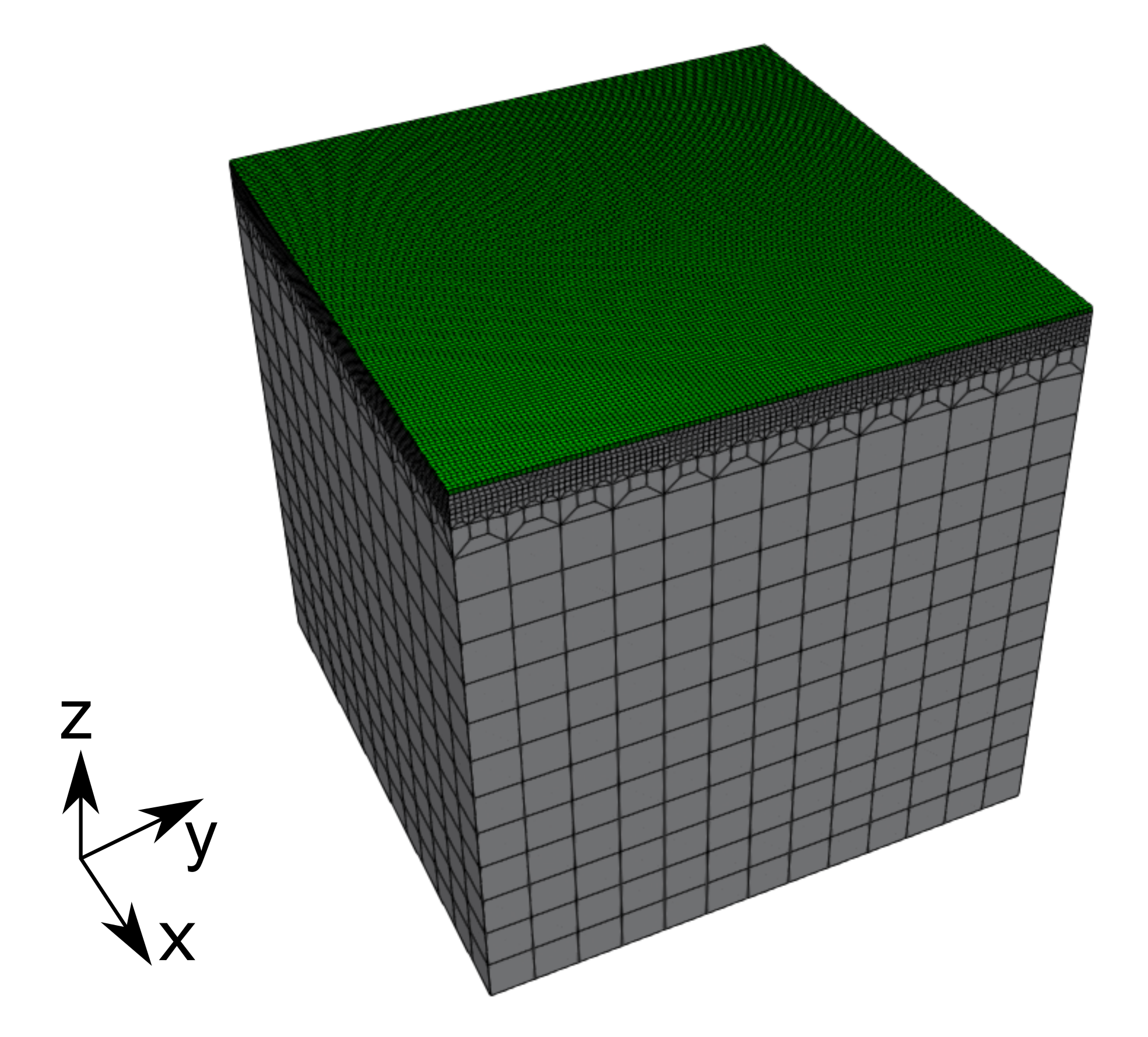}
    \caption{FE model of the benchmark test. The MPJR interface finite elements fully embed the AFM data of the PS-LDPE sample.}
    \label{fig:model}
\end{figure}

The finite element model, shown in \figref{fig:model}, consists of a deformable cubic domain in contact with the layer of the MPJR interface finite elements, which embeds the real composite topography and the adhesive properties of the PS-LDPE sample. 

The cubic domain is discretized using a finer mesh near the contact interface and increasing the size of the finite element mesh at the external part of the domain, as shown in \figref{fig:model}(b). The interface is mounded with $128 \times 128$ MPJR elements, which conform to the solid mesh in the refined region. 

The boundary conditions consist of fixing the degrees of freedom of the upper surface of the interface and applying an upward vertical displacement to the lower surface of the substrate. Finally, no constraints are introduced on the free lateral surfaces of the elastic bulk. 

\begin{figure}[h!]
    \centering    \includegraphics[width=0.3\textwidth]{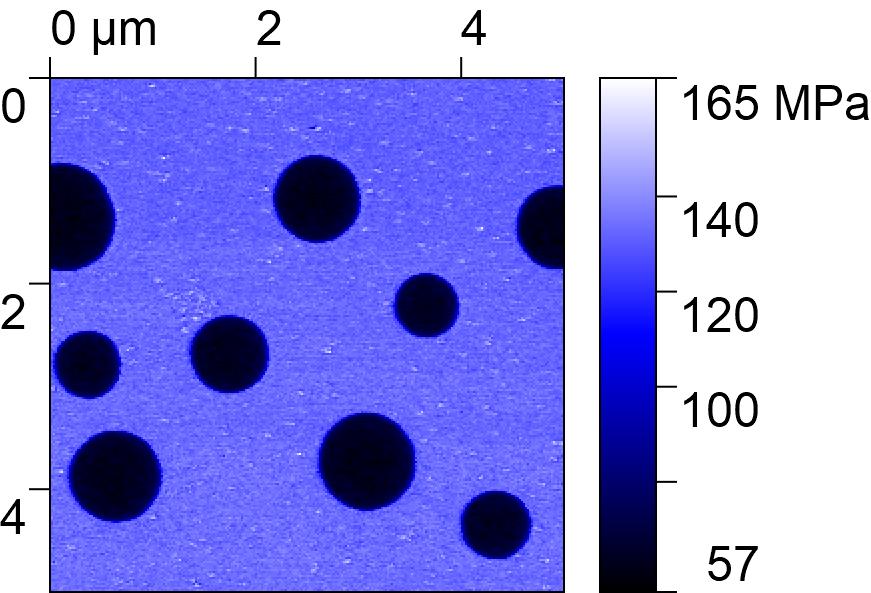}
    \caption{Spatial variation of the DMT elastic modulus of the PS-LDPE sample.}
    \label{fig:logdmt}
\end{figure}

In this first benchmark test, the PS-LDPE sample was considered a linear elastic homogeneous material, assuming an effective elastic modulus value, denoted with $E^*$, calculated from the values measured by the AFM. The experimental data show that the LDPE elastic modulus is lower than that of PS due to the olefin structure in the molecule chain, as shown in \figref{fig:logdmt} in terms of DMT elastic modulus. The average DMT modulus for the two phases has been calculated from the AFM data: $E_{PS}=132.03 \,\si{MPa}$ and $E_{LDPE}=66.88\,\si{MPa}$). The effective value has been obtained from a simple mixture rule using the PS and LDPE volumetric ratios (respectively equal to $0.85$ and $0.15$) as follows:
\begin{equation}\label{eq:vol_ratio}
    E^*=\frac{V_{LDPE}}{V_{LDPE}+V_{PS}}E_{LDPE} + \frac{V_{PS}}{V_{LDPE}+V_{PS}}E_{PS}=121.76\,\si{MPa}
\end{equation}
The volumetric ratios have been estimated assuming that the surface composition given by the AFM scan corresponds to the volume percentage in the bulk. The Poisson ratio is set equal to $\nu=0.32$ as in \citep{Sjoerdsma1981} for a PS-LDPE blend of similar composition.

The simulated quantities, including the heights, adhesion, and dissipated energy used as input for the interface, have been normalized with respect to the lateral size of the sample $L=5\,\si{\mu m}$ and the effective elastic modulus $E^*$. 

\begin{figure}[h!]
    \centering
    \includegraphics[width=0.45\textwidth]{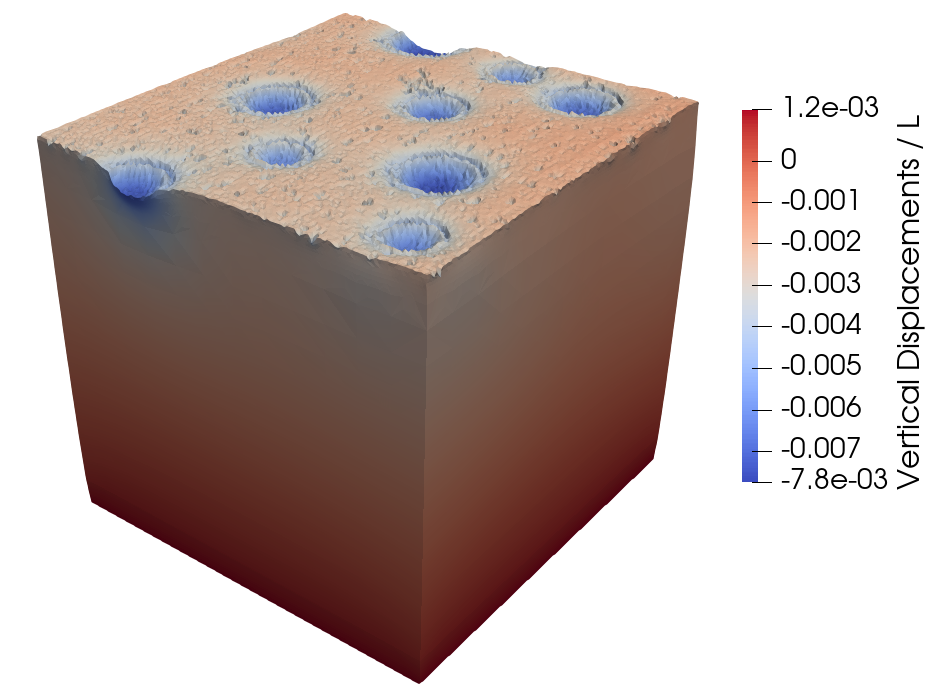}
    \caption{Deformed configuration of the sample (the deformation is magnified $10\times$ for visualization purposes) for a vertical displacement of $-0.25 h_\text{rms}$ applied to the indenter which is pushed against the PS-LDPE sample.}
    \label{fig:displz}
\end{figure}

The simulation results for an applied displacement along the $z$-axis equal to $0.25 h_\text{rms}$ are shown in the contour plot of the vertical displacements $u_z$, \figref{fig:displz}. The effect of the PS-LDPE height field and the L-J repulsive forces can be seen on the indented surface.

\begin{figure}[h!]
    \centering
    \subfloat[Section normal to the axis $y$, at $x=0.38L$]{    \includegraphics[width=0.25\textwidth]{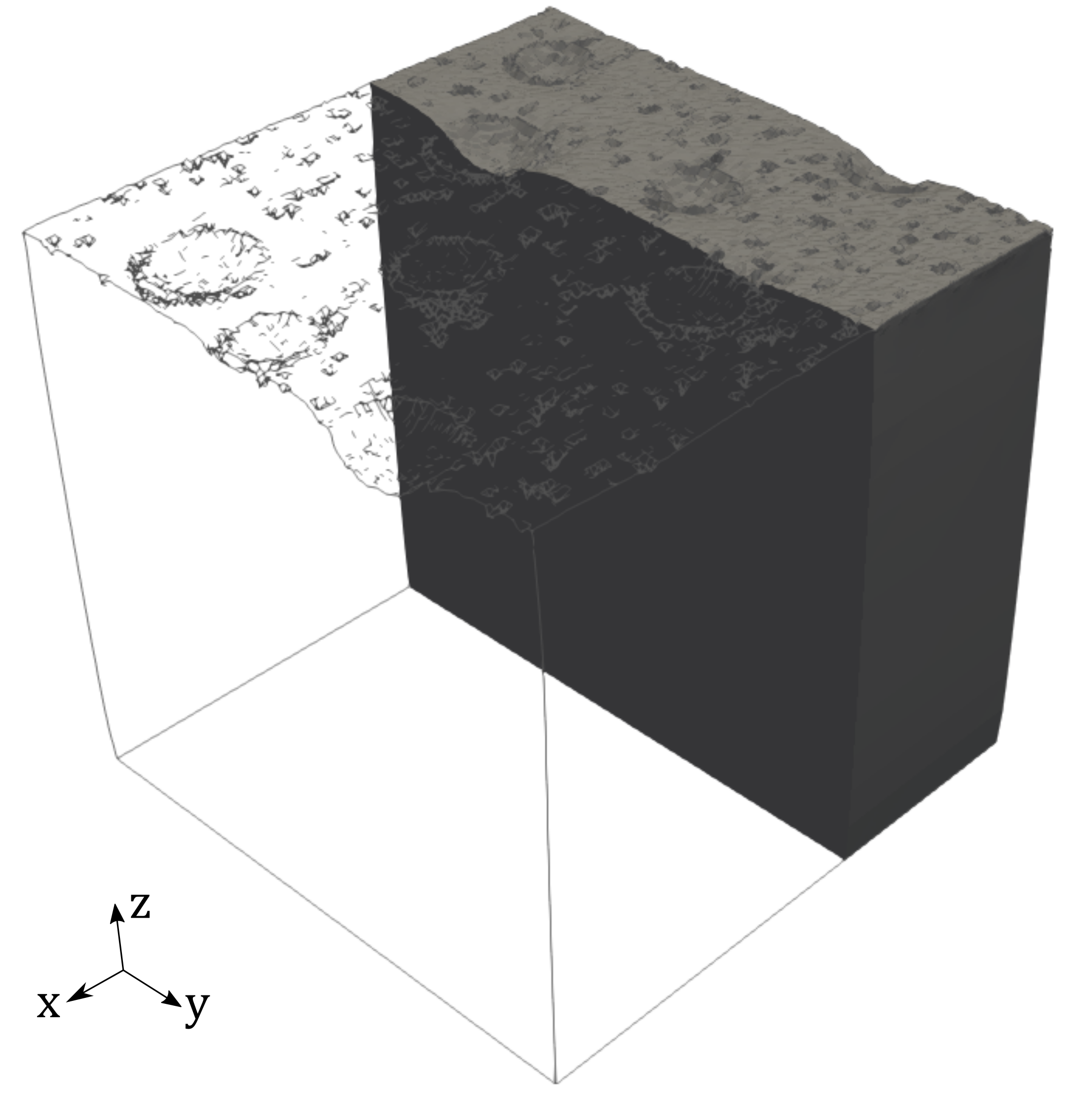} \qquad \qquad
    \includegraphics[width=0.6\textwidth]{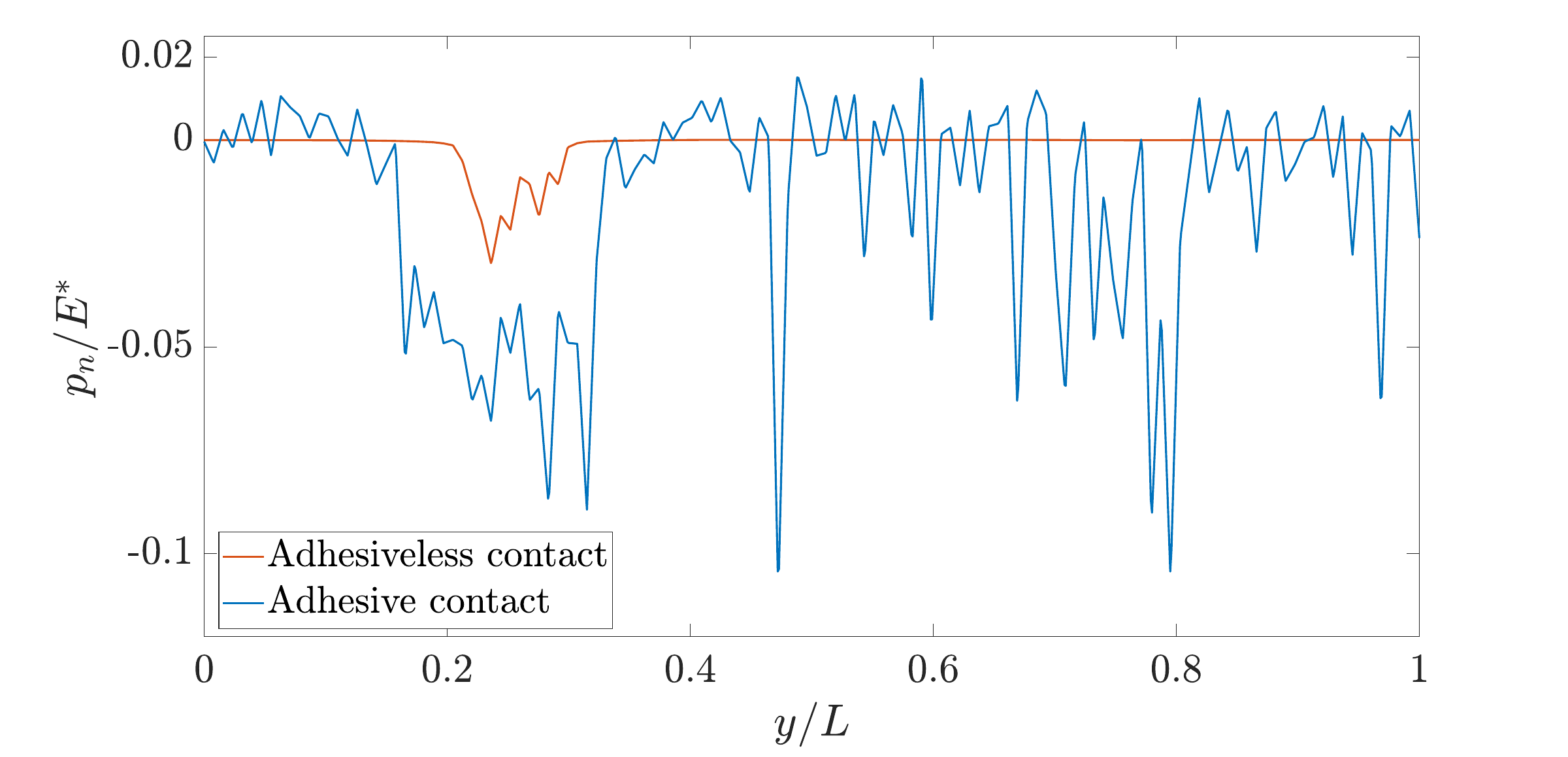}}\\
\subfloat[Section normal to the axis $x$, at $y=0.75L$]{          
    \includegraphics[width=0.25\textwidth]{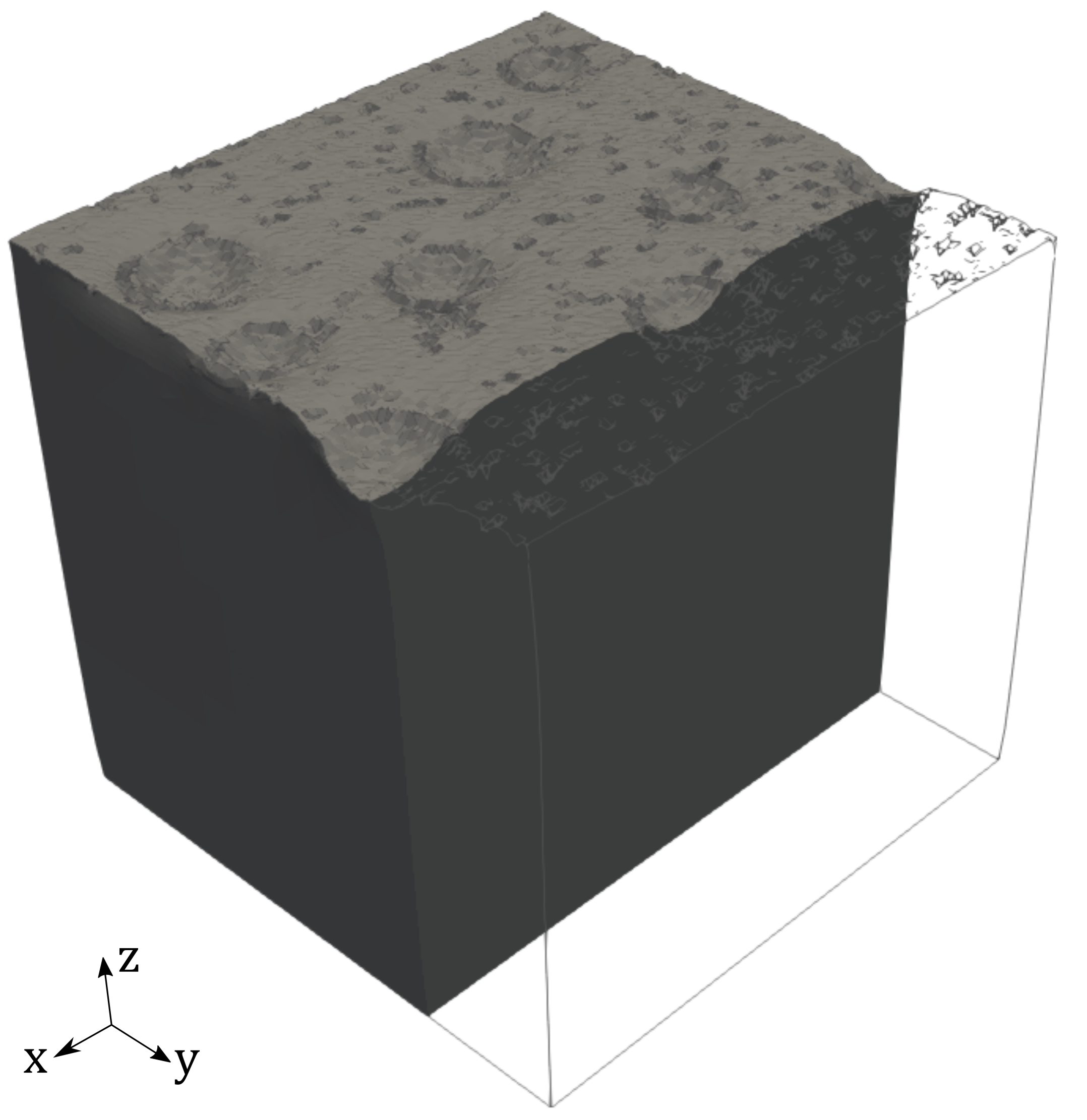} \qquad \qquad
    \includegraphics[width=0.6\textwidth]{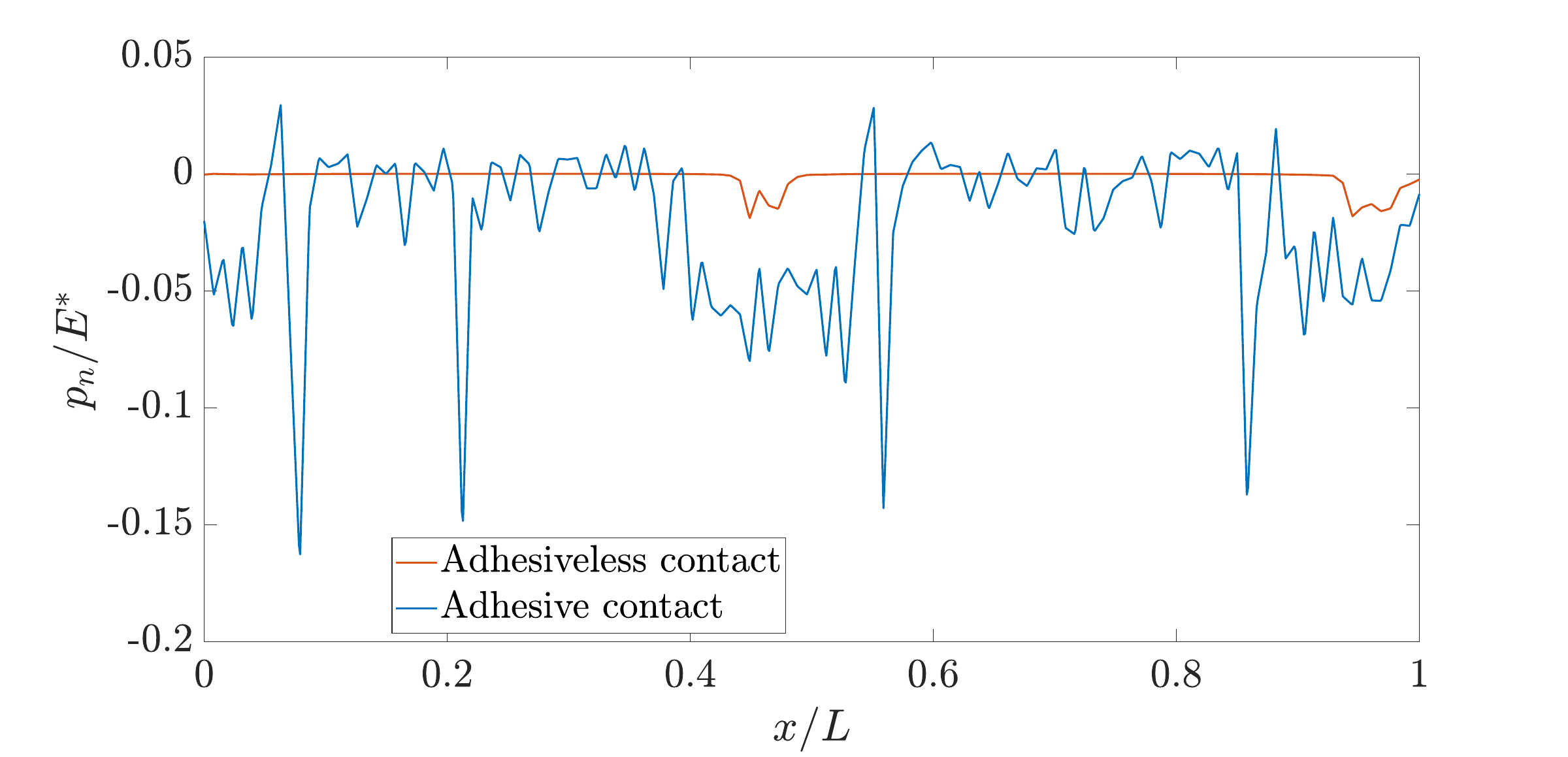}}
    \caption{Comparison of the dimensionless normal traction  $\sigma_n/E^*$ at two chosen interface sections, with or without the adhesive law. Imposed far-field displacement $\bar{u}=-0.25h_\text{rms}$.}
    \label{fig:sigma}
\end{figure}

The results obtained with the described approach have been compared with those obtained by neglecting the adhesion contribution, considering a penalty approach with stiffness set equal to the average penalty factor used for the adhesive case. The comparison of the resulting dimensionless normal traction is shown in \figref{fig:sigma} for two interface sections in directions $x$ and $y$. The model considering the Lennard-Jones potential is characterized by regions having positive tractions (adhesion) and higher negative tractions caused by the repulsive forces between the surfaces. The standard penalty approach does not consider the equilibrium distance between the solids.

\begin{figure}[h]
    \centering
    \subfloat[Section normal to the axis $y$, at $x=0.5L$]{\includegraphics[width=0.3\textwidth]{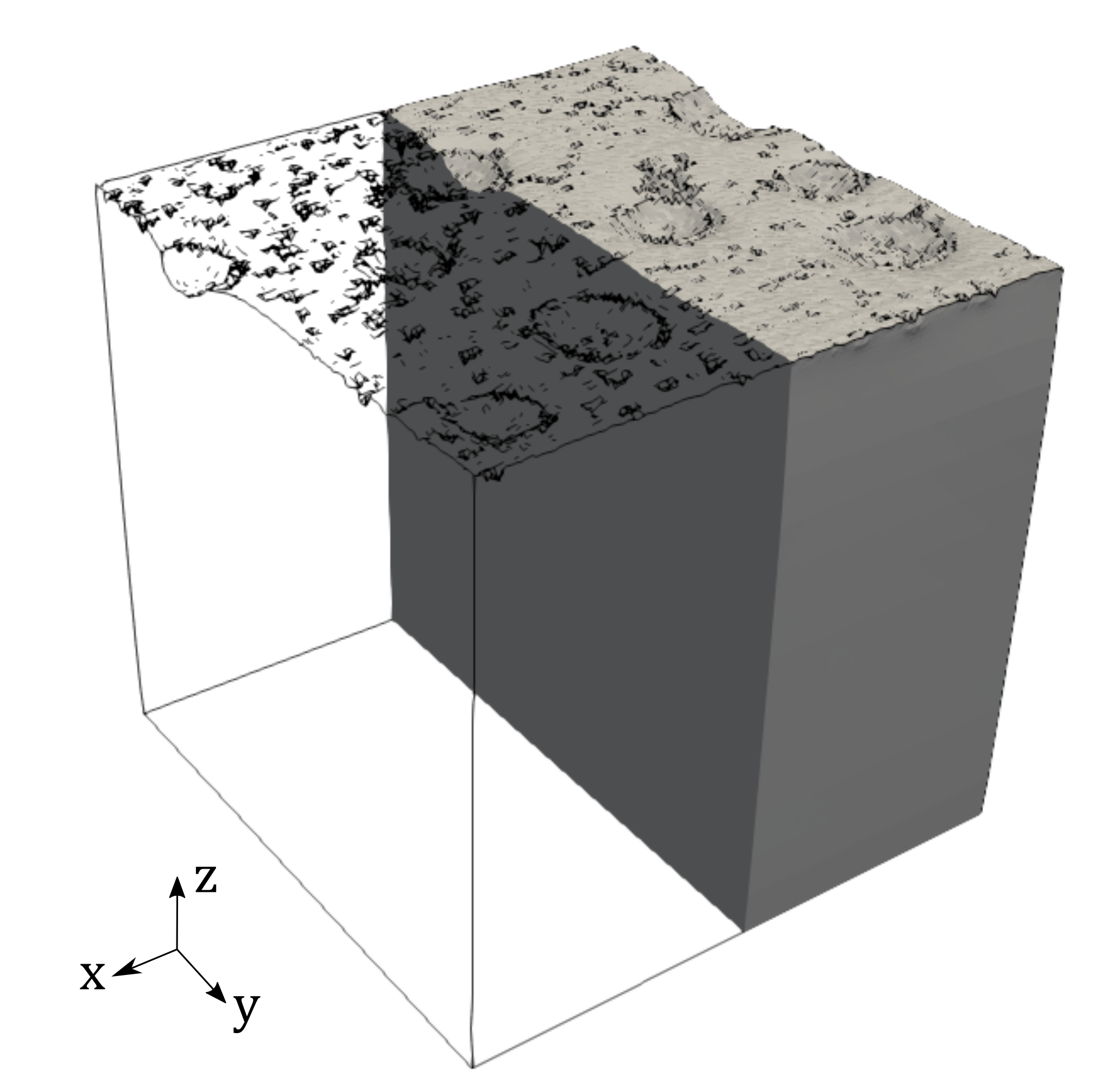} \qquad 
    \qquad    \includegraphics[width=0.6\textwidth]{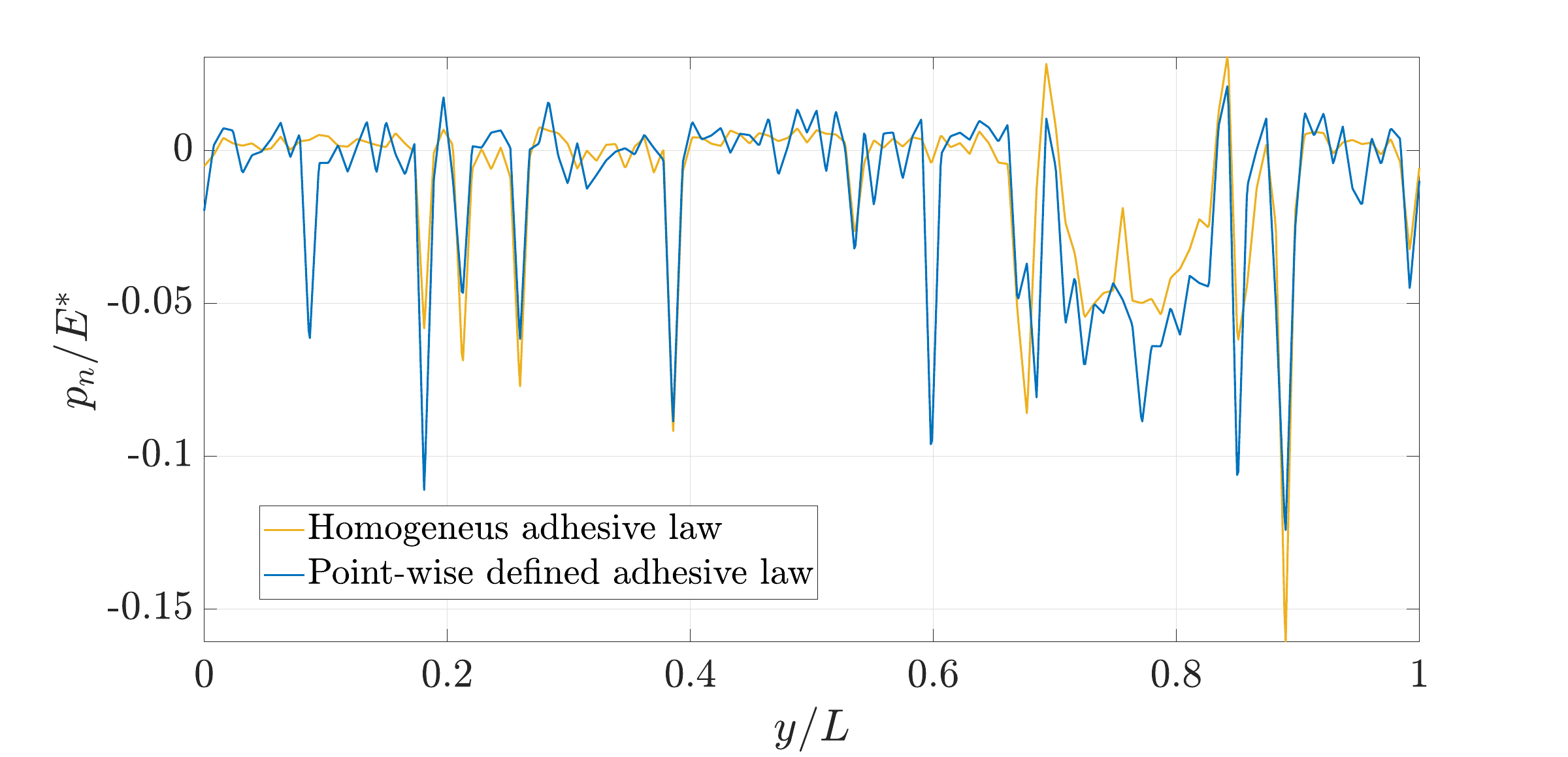}}\\
    \subfloat[Section normal to the axis $x$, at $y=0.5L$]{\includegraphics[width=0.3\textwidth]{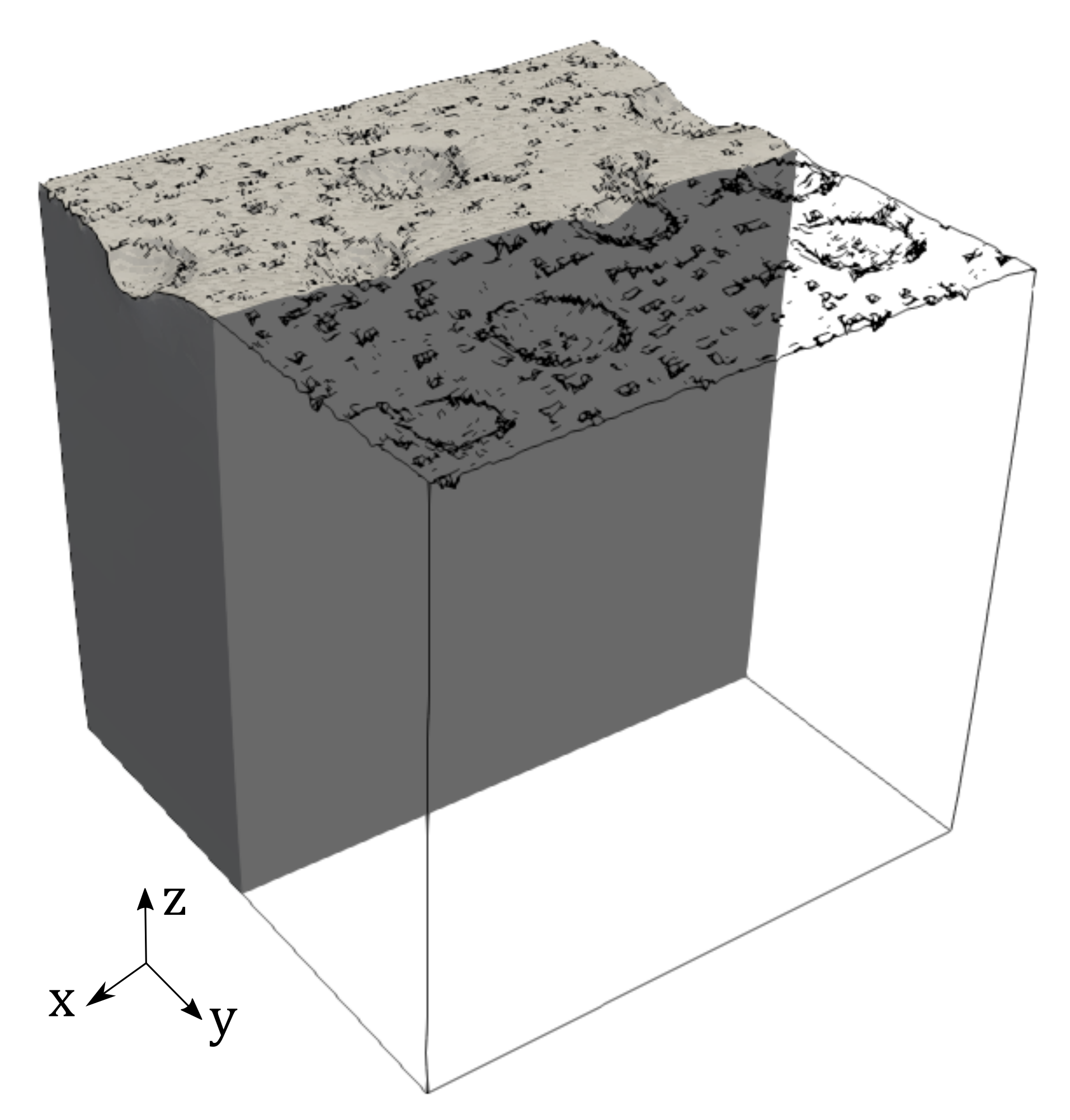} \qquad 
    \qquad    \includegraphics[width=0.6\textwidth]{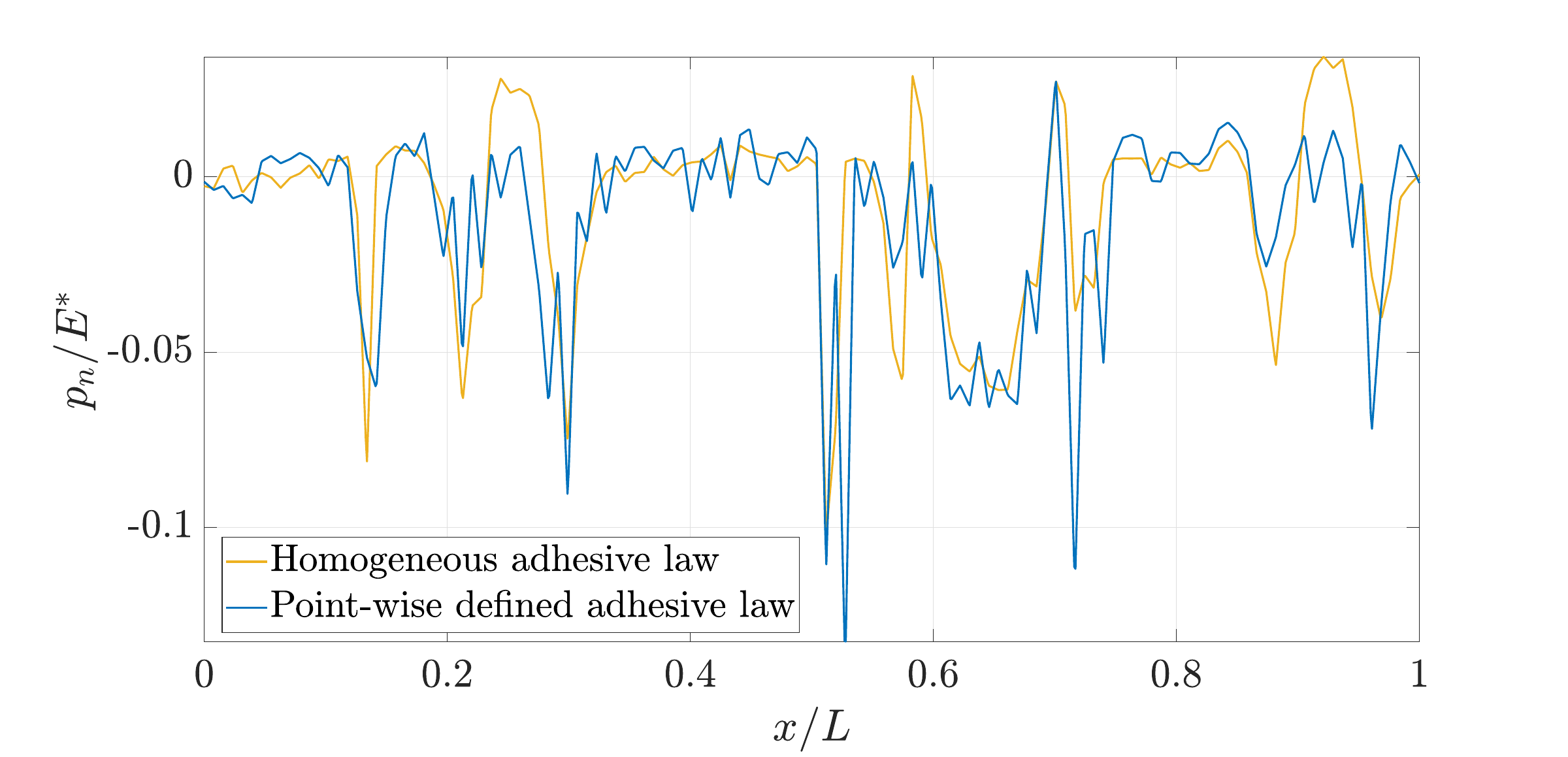}}
    \caption{Comparison of  $\sigma_n/E^*$ at two chosen interface sections for the model having the coefficients of the interface constitutive law defined point-wise according to the ATM data, versus the the model using homogeneous, averaged coefficients. Imposed far-field displacement $\bar{u}_z=0.2h_\text{rms}$ (the indenter is pull-out from the sample).}
    \label{fig:sigma_adh}
\end{figure}

As previously described, the proposed modeling approach includes an interface constitutive law point-wise defined according to the field data collected with the AFM. This aspect is highlighted in \figref{fig:sigma_adh} where the simulation results have been compared with the same simulation considering a homogeneous constitutive law whose parameters for the Lennard-Jones relationship are set equal to the average values of the parameters provided by the experimental results. In this case, the results are plotted for the imposed displacement $\bar{u}_z=0.2 h_{rms}$, corresponding to moving away the indenter from the substrate. While in the homogenized case, the contact stress variation depends only on the surface roughness, the point-wise defined constitutive law allows the simulation of the combined effect of adhesion and roughness. 

\clearpage
\section{Assessing the role of the bulk heterogeneity}\label{sec:elastic_mod}

In the previous section, the sample was considered as a homogeneous linear elastic material with an effective (homogenized) elastic modulus and heterogeneous interface properties. In this section, we remove the simplification of a homogenized continuum, and we investigate the effect on the contact response of using dissimilar elastic properties in the bulk as a closer representation of the real internal heterogeneity. Hence, different elastic moduli for the two polymeric phases of the PS-LDPE sample are considered.

\begin{figure} [h]
    \centering
    \includegraphics[width=0.3\textwidth]{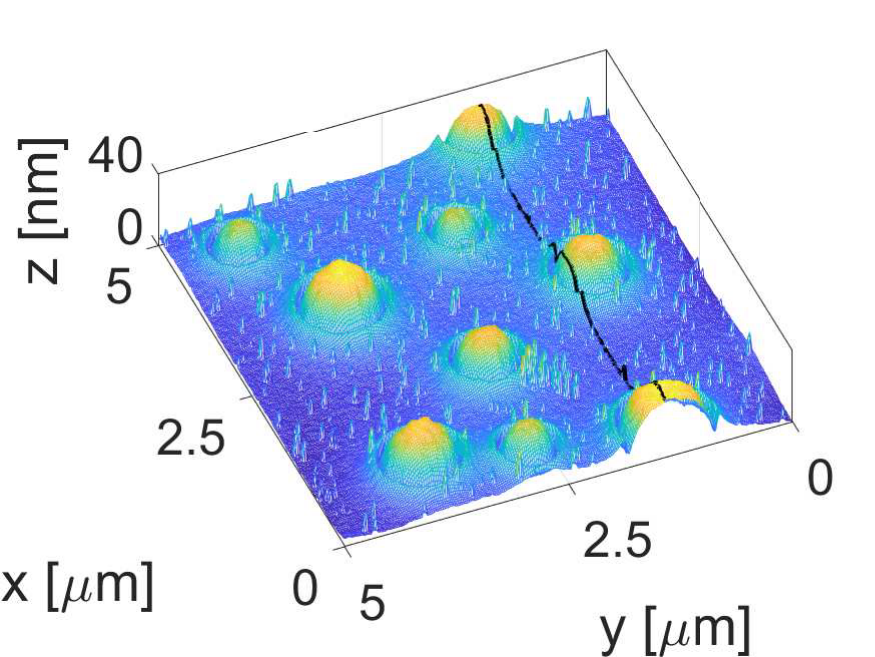}\qquad
    \includegraphics[width=0.5\textwidth]{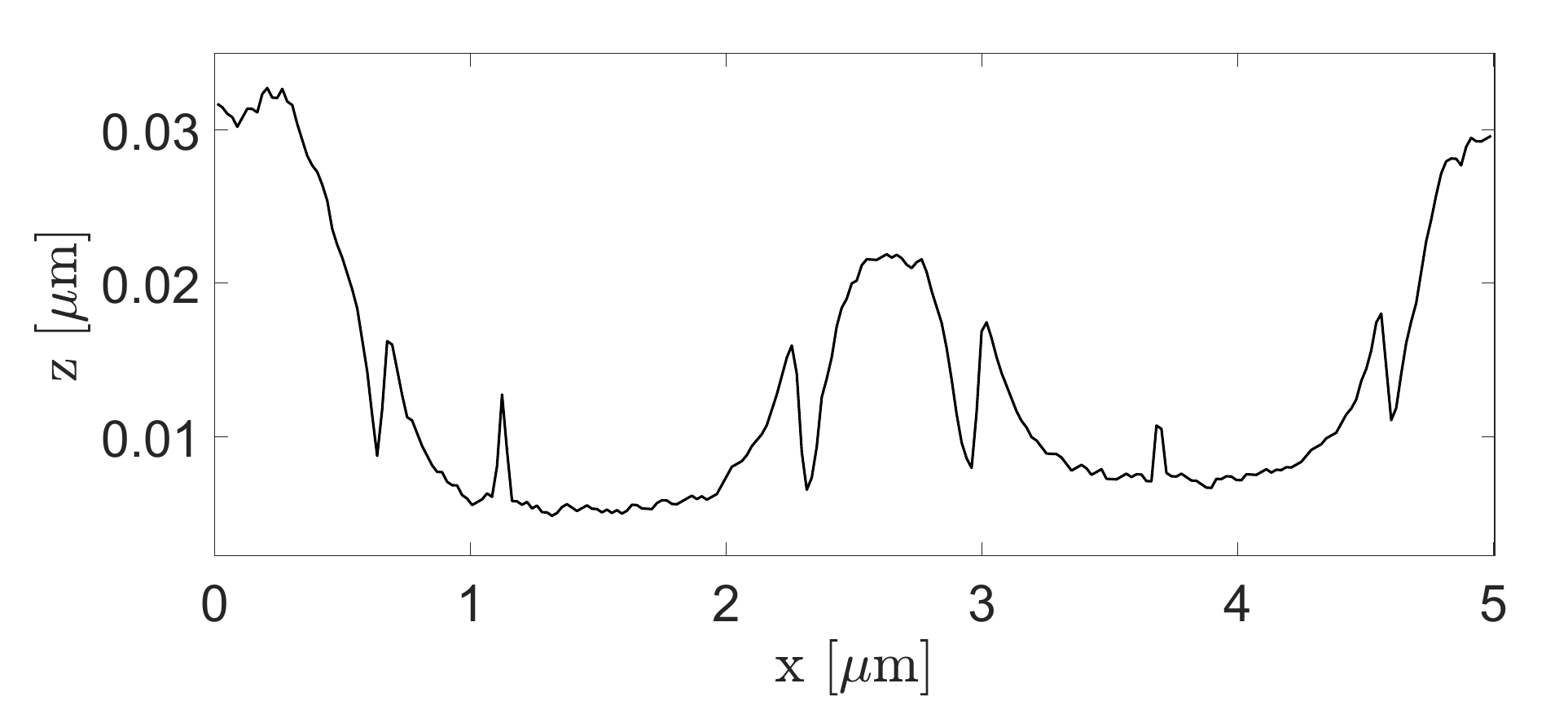}
    \caption{PS-LDPE profile extracted from the surface topography, for the 2D simulations.}
    \label{fig:profile}
\end{figure}

For this simulation, a single rough profile has been extracted from the surface topology as shown in \figref{fig:profile} and imported in the 2D interface finite elements, together with the corresponding values of maximum adhesion force and energy dissipation at each profile coordinate. 

\begin{figure}[h]
    \centering
    \includegraphics[width=0.9\textwidth]{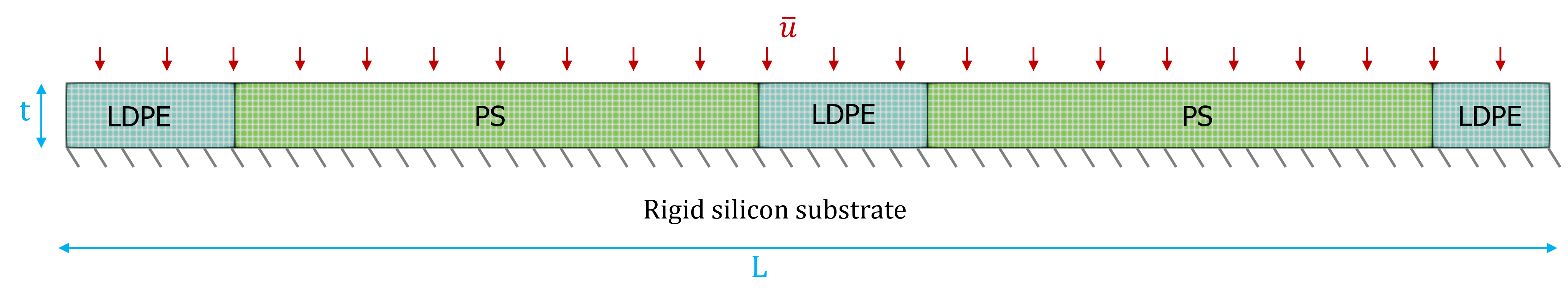}
    \caption{2D geometry and boundary conditions of the sample showing the two polymeric phases, PS and LDPE. The sample thickness is $t=0.2\tild\si{\mu m}$, and the length is $L=5\tild\si{\mu m}$.}
    \label{fig:geometry2d}
\end{figure}

The geometry of the FE model is depicted in \figref{fig:geometry2d}. Compared to the 3D model of the previous section, the 2D model thickness $t$ has been set equal to $0.2\tild\si{\mu m}$, considering that the PS-LDPE layer is deposited on a silicon substrate, which can be considered rigid because its Young's modulus is $\approx 150\tild\si{GPa}$, much higher than those of the polymeric components. For this reason, the discretization of the silicon substrate layer has been neglected, and the polymeric layer has been constrained at the bottom edge.

\begin{figure}[h!]
    \centering
    \includegraphics[width=0.7\textwidth]{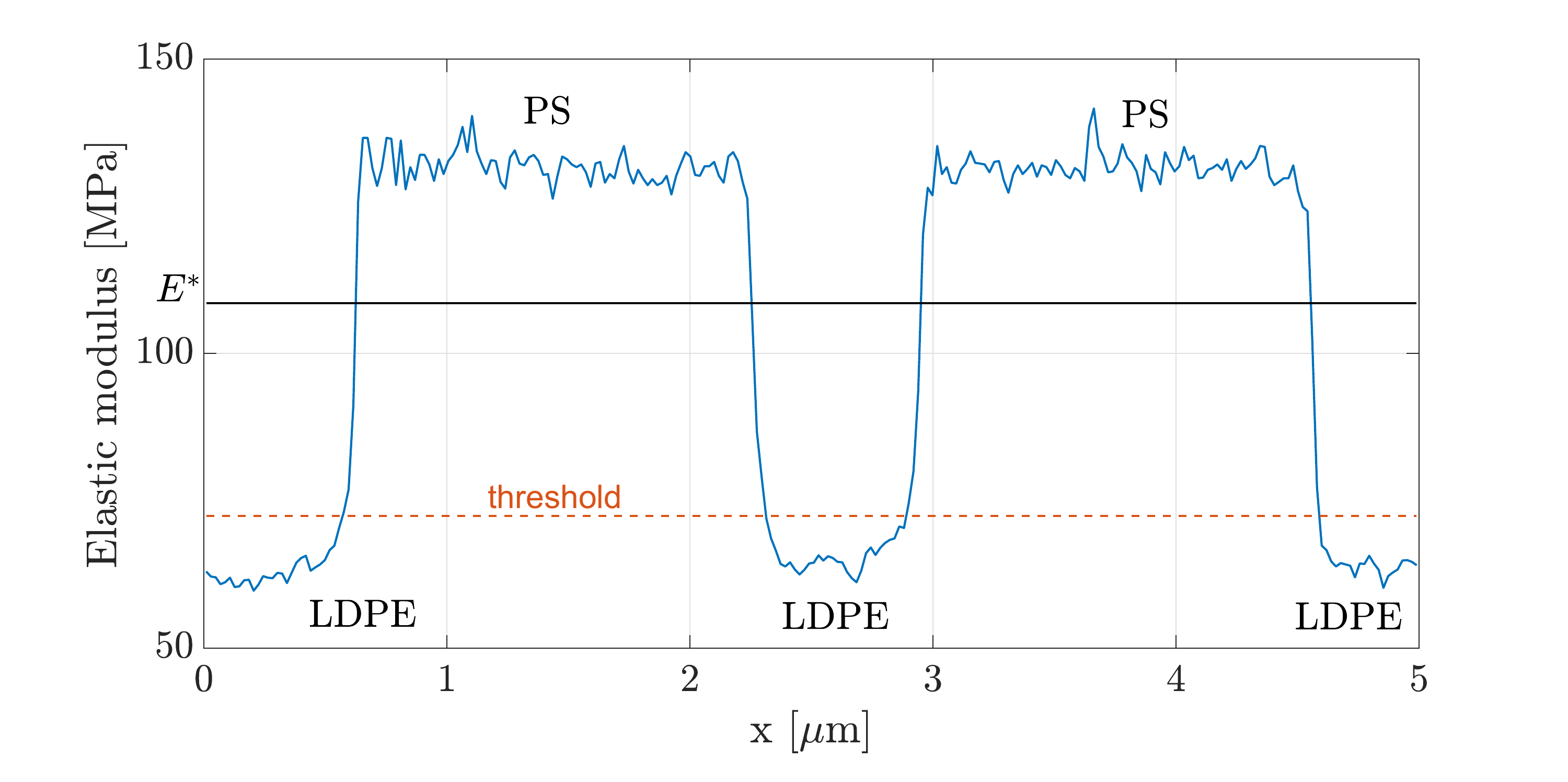}
    \caption{Elastic modulus variation along the analyzed profile. The red dashed line shows the threshold chosen to separate the PS and the LDPE phases; the black continuum line corresponds to the value of effective elastic modulus $E^*=108.5\tild\si{MPa}$ used in the homogenized model.}
    \label{fig:elastic_modulus}
\end{figure}

Figure \ref{fig:geometry2d} also shows how the two polymeric phases alternate in the sample. This variation can be recognized by looking at the elastic modulus values measured with the AFM along the chosen profile and depicted in \figref{fig:elastic_modulus}.
The threshold value of $72.44\tild\si{MPa}$ (represented as a red line in the figure) has been used to distinguish between the two components and the two polymers have the following averaged values of the elastic modulus: $(i)$ $E_1= 128.67\tild\si{MPa}$  for the PS material model, $(ii)$ $E_2= 64.27\tild\si{MPa}$ for the LDPE. 

\begin{figure}[h]
    \centering
    \includegraphics[width=0.7\textwidth]{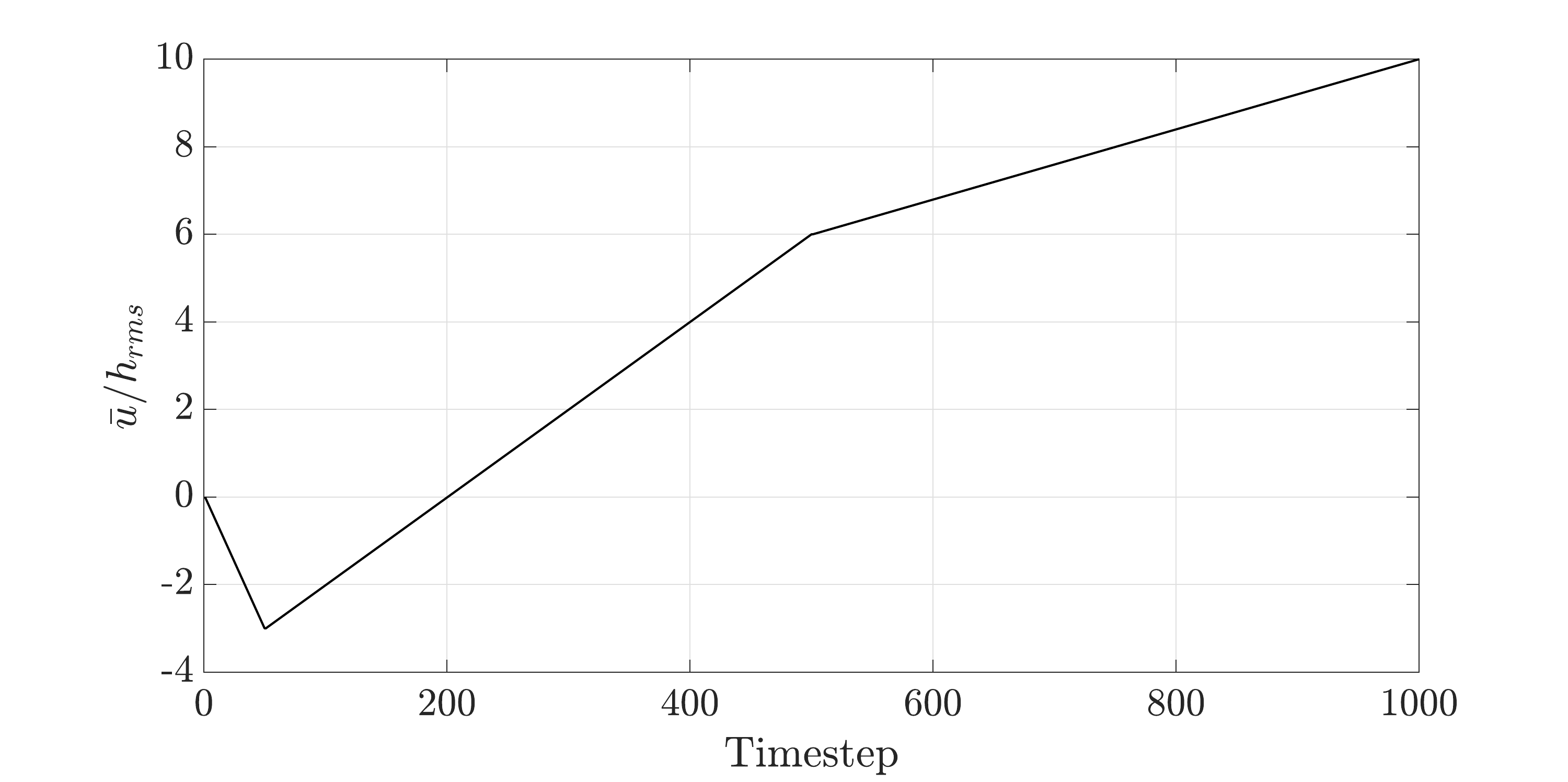}
    \caption{Imposed far-field displacement $\bar{u}$ applied to the top edge of the indenter with respect to a pseudo-time variable. After the first approaching stage, the indenter is pulled out from the substrate.}
    \label{fig:appl_displ}
\end{figure}

The simulation has been conducted by prescribing a non-monotonic far-field displacement path to the indenter, composed of two ramps throughout the pseudo-time evolution. In the first ramp, (denominated approaching ramp), the rough indenter is pushed against the polymeric layer, increasing the imposed far-field displacement $\bar{u}$ up to $\bar{u}=-3 h_\text{rms}$. Then, the far-field displacement is reverted until the two surfaces are pulled apart (denoted as separation ramp). The displacement path is represented in \figref{fig:appl_displ}. 

\begin{figure}[h]
    \centering
    \subfloat[Approach ramp.]{    \includegraphics[width=0.7\textwidth]{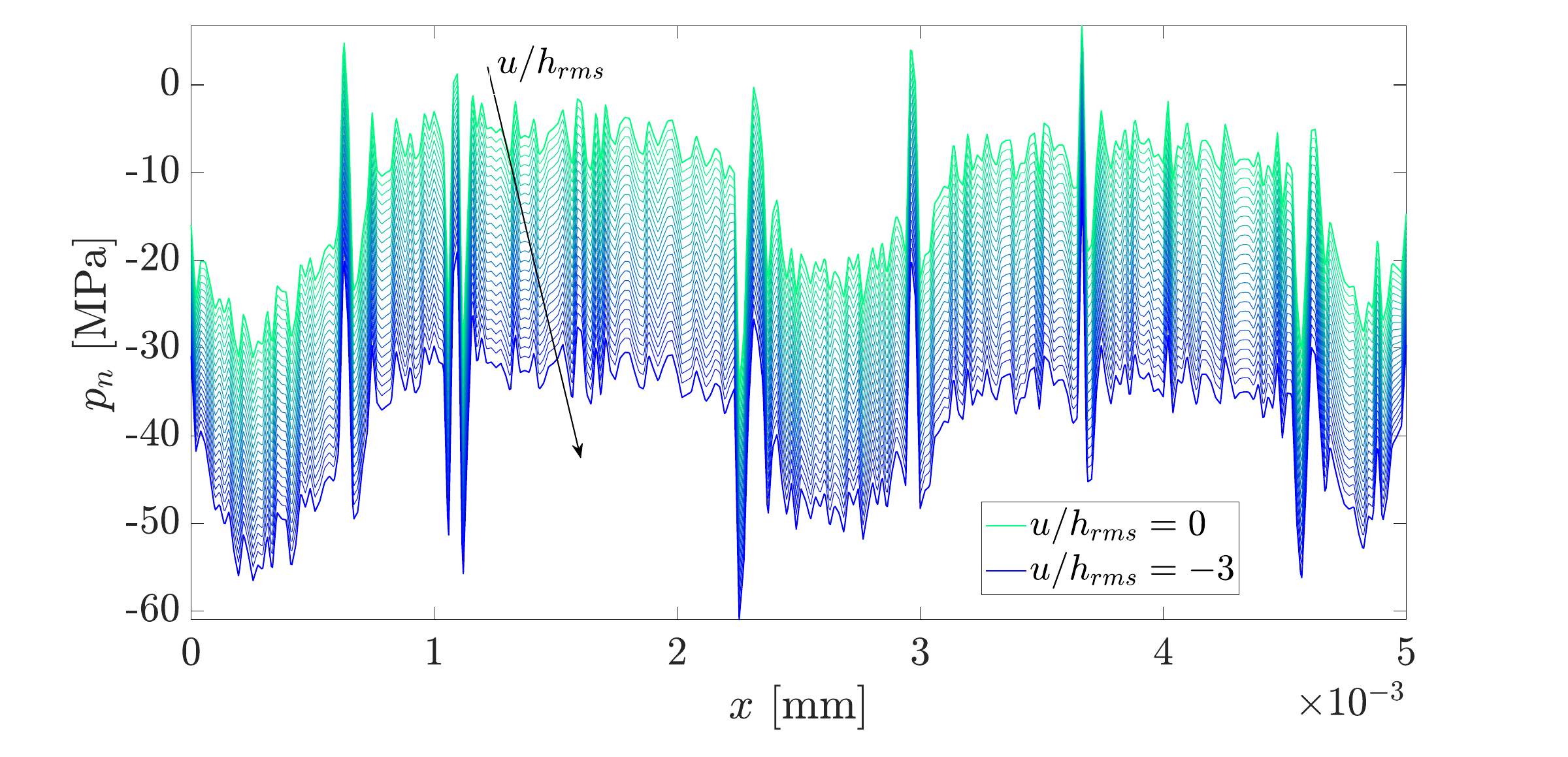}}\\
    \subfloat[Separation stage till the adhesive peak force.]{    \includegraphics[width=0.7\textwidth]{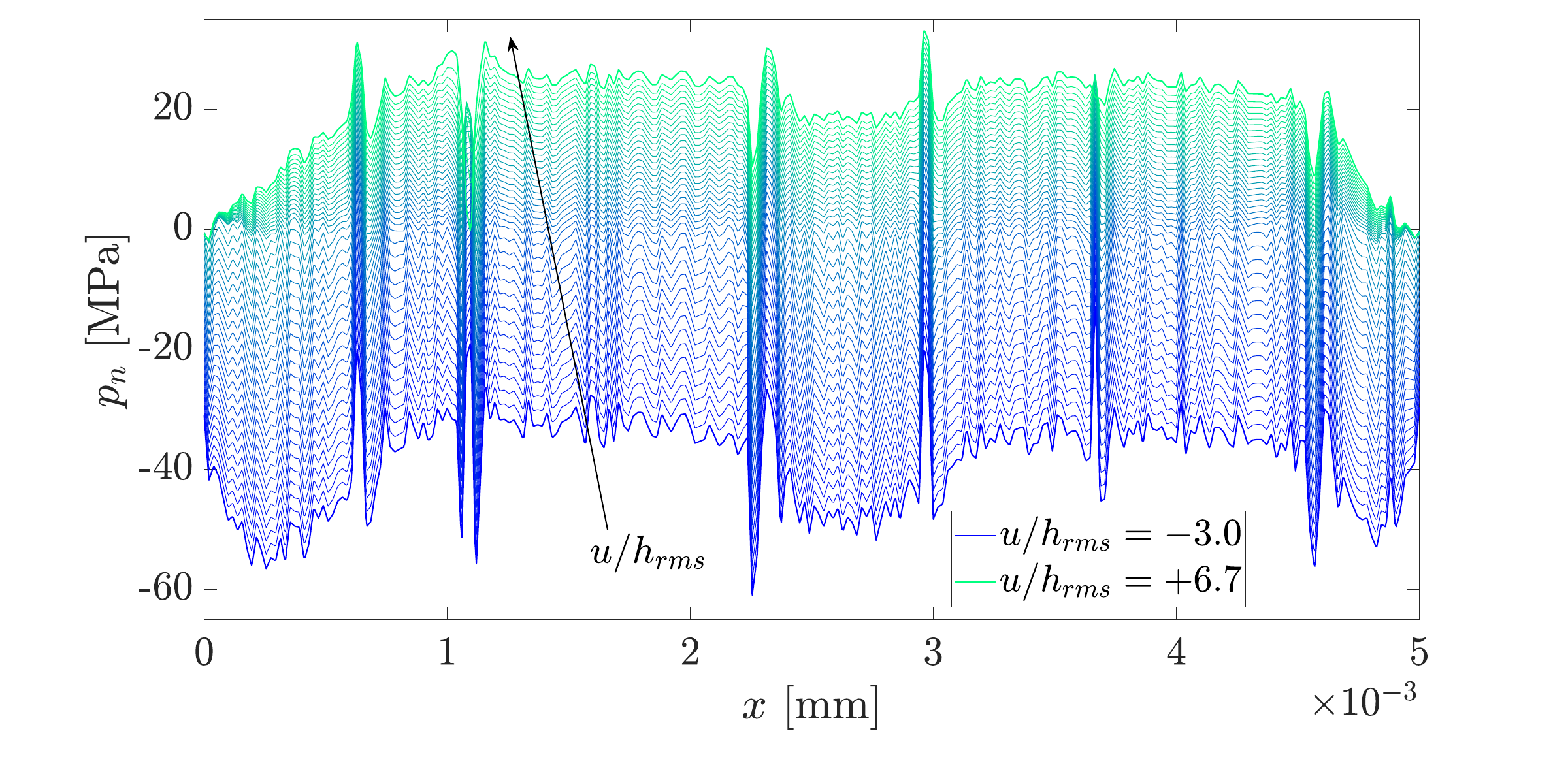}}\\
    \subfloat[Separation stage from the adhesive peak force to the almost complete detachment of the indenter.]{    \includegraphics[width=0.7\textwidth]{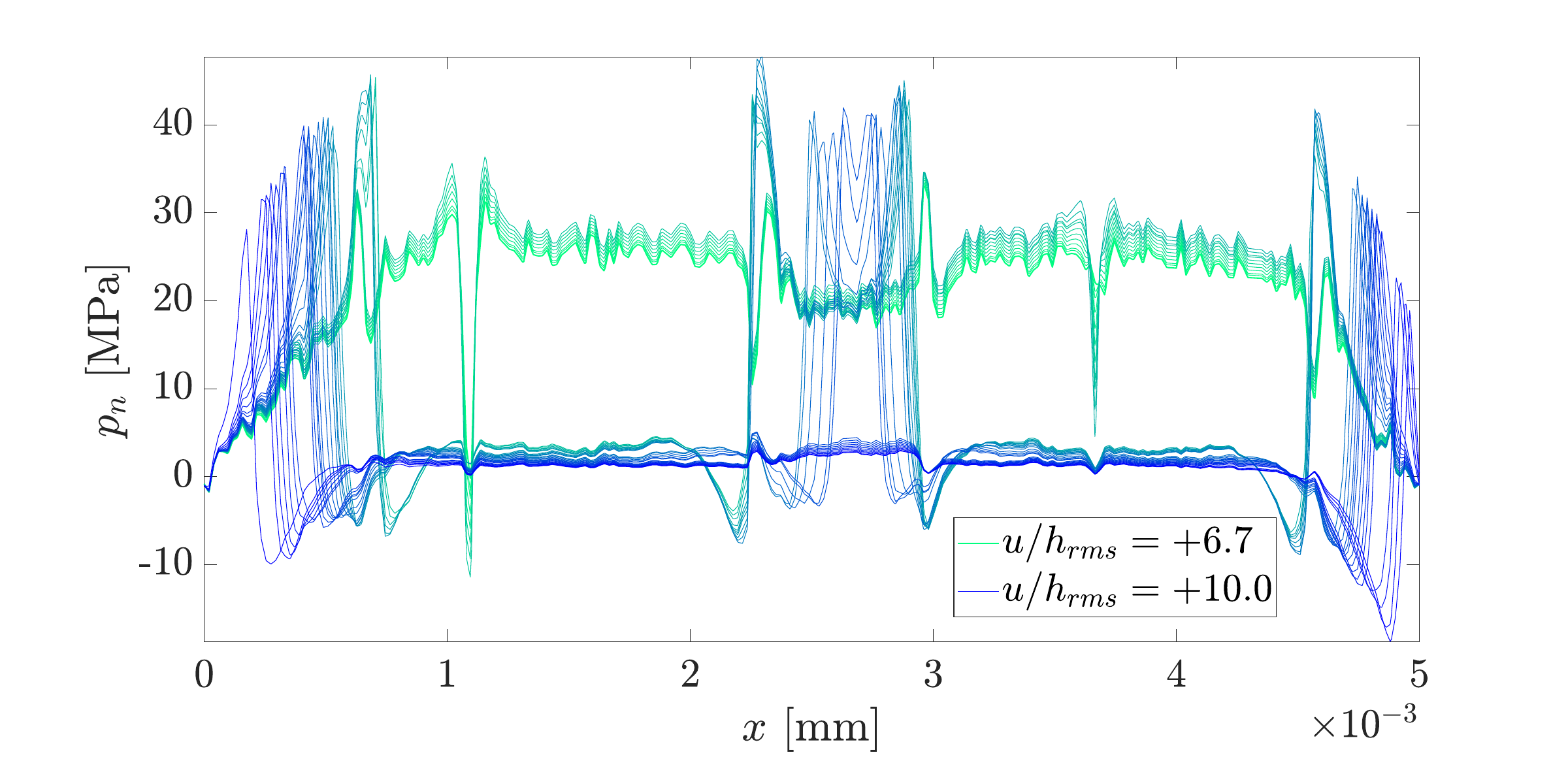}}
    \caption{Contact tractions variation during the approach ramp in (a) and the following separation stage in (b-c). The total reaction force increases till the maximum adhesive force, corresponding to an imposed far-field displacement $\bar{u}=6.7 h_\text{rms}$ in (b), and decreases to almost zero in (c).}
    \label{fig:pull_off}
\end{figure}

Figure \figref{fig:pull_off} depicts the normal contact tractions during the simulation. For visualization purposes, the interface response has been represented in three separate plots: \figref{fig:pull_off}a corresponds to the approach ramp;  \figref{fig:pull_off}b shows the contact tractions from the beginning of the separation ramp to the time-step when the maximum adhesive force is achieved (corresponding to $\bar{u}=6.7 h_\text{rms}$ and $P=0.24 \si{N}$); \figref{fig:pull_off}c represents the decrease in the contact tractions from the maximum adhesive force to almost complete separation of the two surfaces, showing that detachment does not occur simultaneously along the interface. First, it happens in the PS regions and later, the LDPE inclusions also detach from the indenter, which is consistent with the adhesive data collected by the AFM (see \figref{fig:adh}) where the PS matrix has, on average, a lower value of the maximum adhesive force.

\begin{figure}[h]
    \centering
    \includegraphics[width=0.7\textwidth]{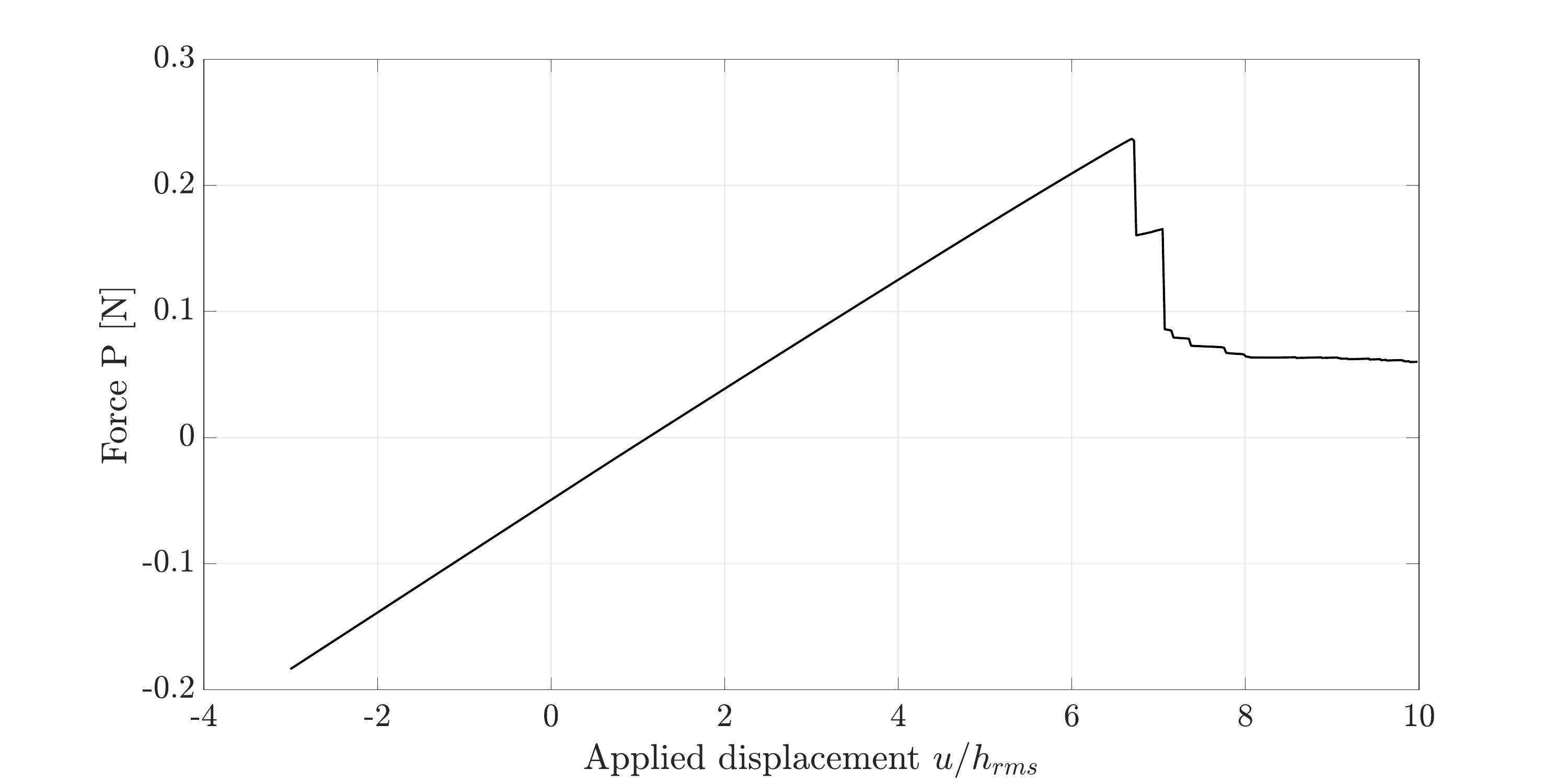}
    \caption{Total reaction force vs. the imposed far-field displacement for the 2D model with the heterogeneous substrate, alternating the PS and the LDPE phases as in the real sample.}
    \label{fig:reac}
\end{figure}

The total resulting reaction force is plotted in \figref{fig:reac} with respect to the applied displacement. As expected, the reaction force during the approaching ramp overlaps with the initial part of the separation ramp since the energy dissipation during the approach stage has been neglected when computing the interface constitutive law. Moreover, the total reaction force shows a snapback instability after reaching the maximum adhesive force. 


The instabilities related to adhesive contact problems have been intensively studied in the literature (see, e.g., \cite{guduru1, guduru2, Ciavarella2019,SANNER2022} and references therein). The traction law in \eqref{eq:len-jo} can be regarded as a nonlinear spring in series with the linear spring associated with the deformable substrate (assuming a rigid indenter) having stiffness $k$, which can be assumed equal to $E/t$, where $E$ is the elastic modulus and $t$ is the substrate depth in the undeformed configuration. This phenomenon is analogue to the case of a nonlinear interface law with softening behavior studied in \citep{Carpinteri1989}. That work can be used as a guideline to identify the occurrence of instabilities for the proposed Lennard-Jones-like constitutive law exploited in the present work. In fact, instabilities can appear when:
\begin{equation}\label{eq:inequalities}
   \left|\frac{\partial p_n}{ \partial g_n}\right|_{\max} > \frac{E}{t} 
\end{equation}
where $\left|\frac{\partial p_n}{ \partial g_n}\right|_{\max}$ is the maximum module of the tangent of the interface constitutive law. Considering the average values of the interface parameters used for the PS-LDPE example, the tangent maximum value is $|\partial p_n/\partial g_n|_{\max}=2.17 \times 10^6$ N/mm$^3$. Using $t=0.0002\,\si{mm}$ (substrate depth) and $E=108.5 \,\si{MPa}$, we obtain $E/t=5.427 \times 10^5 \,\si{N/mm^3}$. Hence, the present situation satisfies the inequality in \eqref{eq:inequalities}, which confirms the occurrence of instabilities revealed in the reaction force-displacement curve. 

\begin{figure}[h]
    \centering
    \includegraphics[width=0.7\linewidth]{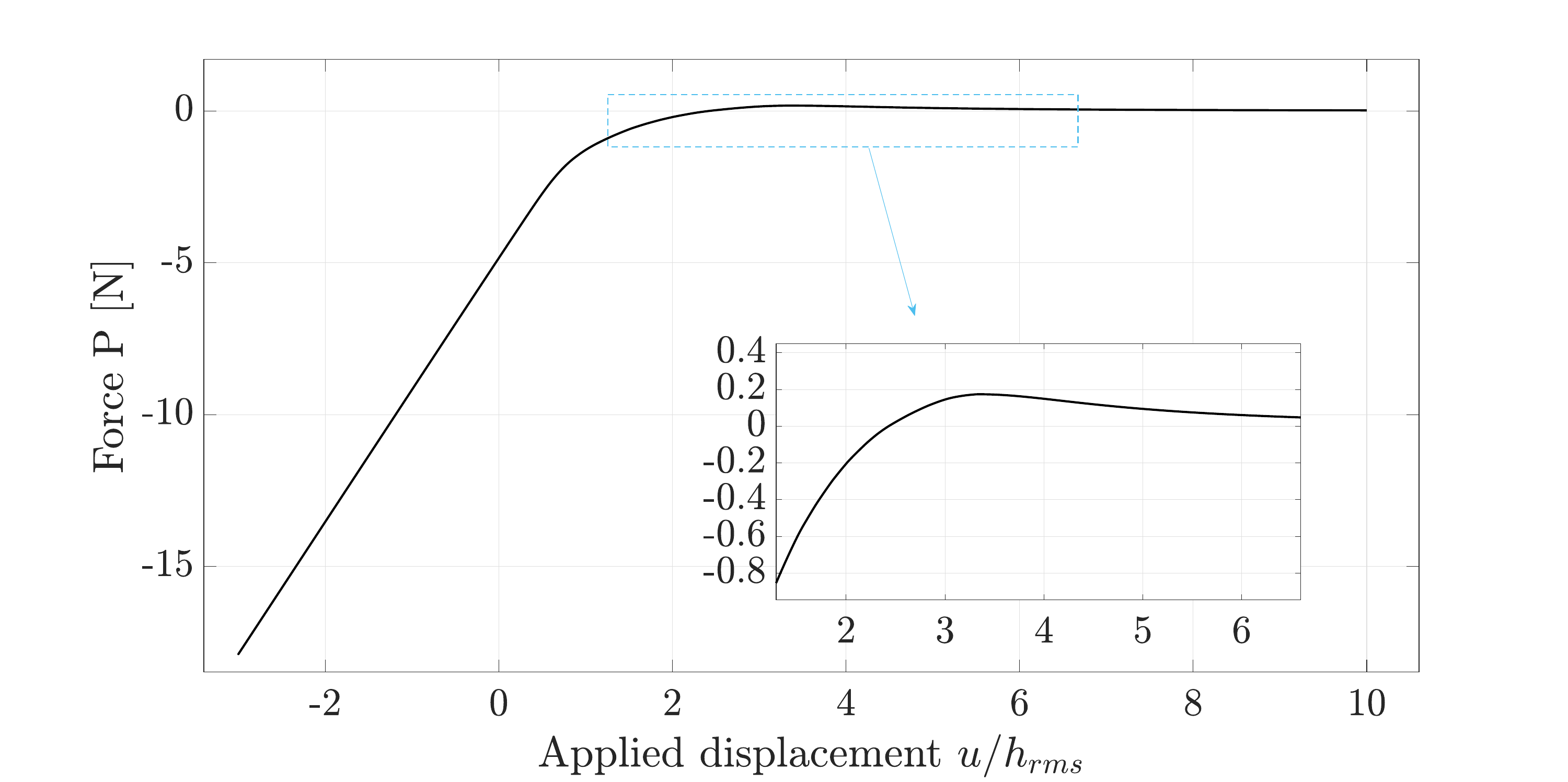}
    \caption{Total reaction force vs. applied displacement for an exemplary case with elastic moduli two orders of magnitude bigger than in Fig.~\ref{fig:reac}. This case does not show snap-back instabilities.}
    \label{fig:force_displ_higher_E}
\end{figure}

The present case can be compared with another hypothetical situation where the interface properties are the same, while the elastic moduli of the two components are assumed to be two orders of magnitude higher than before: $(i)$ $E_1= 12867\, \si{MPa}$ for the first phase, $(ii)$ $E_2= 6427\, \si{MPa}$ for second phase. For this case, the above inequality is not satisfied, since $E/t=5.427\times 10^7 \,\si{N/mm^3}$. This is confirmed in the simulation results plotted in \figref{fig:force_displ_higher_E} where no instabilities are observed. This is also confirmed by the value of the Tabor parameter \citep{CIAVARELLA2019b,Ciavarella2019} associated with the two cases. The Tabor parameter, initially proposed for the adhesive contact between a sphere and an elastic half-plane, is inversely proportional to the elastic modulus. Hence, for lower values of the elastic modulus, the simulated contact problem tends to behave according to the Johnson–Kendall–Roberts (JKR) model \citep{Johnson1971} characterized by a \emph{S-shaped} force-displacement curve.

In order to demonstrate the versatility of the MPJR approach for heterogeneous contact problems, the present model results have been compared to the homogenized simulation, where the elastic modulus of the bulk is set equal to the effective (homogenized) value $108.5\tild\si{MPa}$ for the PS and LDPE phases, computed according to Eq.~\eqref{eq:vol_ratio} based on the volumetric ratios selected for the chosen profile.

\begin{figure}[h]
    \centering
    \includegraphics[width=0.6\textwidth]{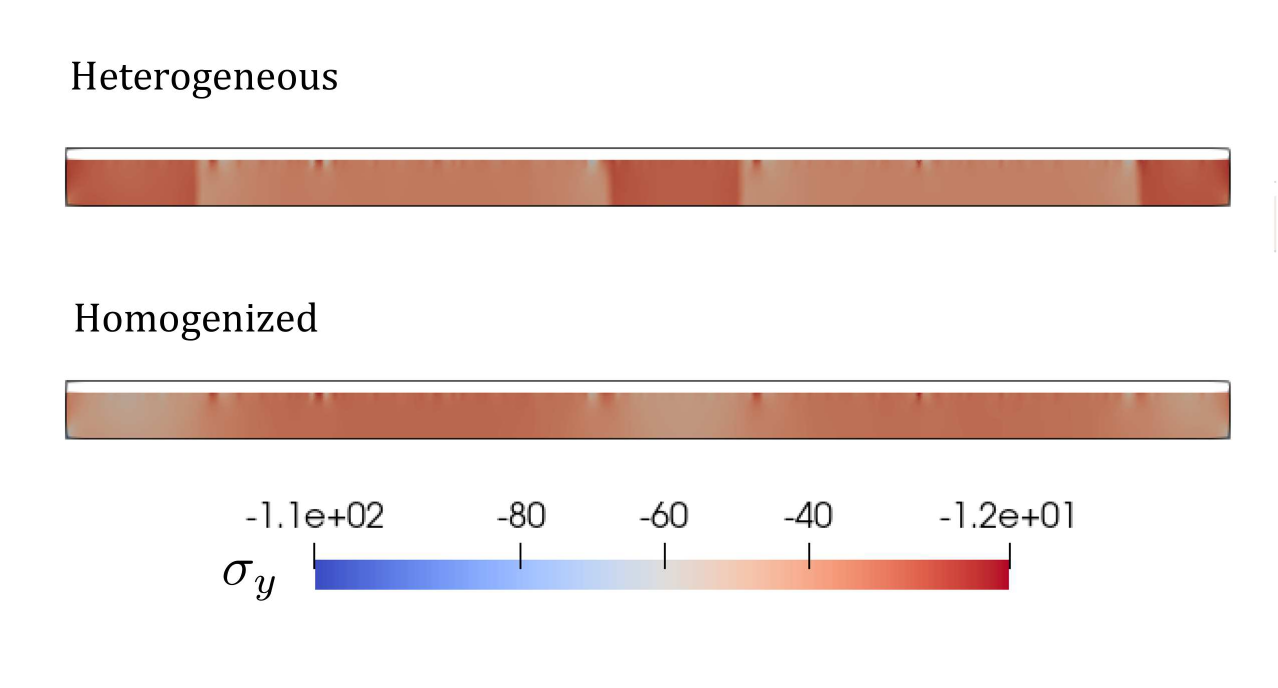}
    \caption{Contour plot of $\sigma_y$ resulting from the homogenized model having elastic modulus $E^*$ and in the case of the model with two different elastic modulus values for the PS and LDPE phases, for an imposed displacement of $\bar{u}=-3 h_{rms}$.}
    \label{fig:stress_domain}
\end{figure}

\begin{figure}[h]
    \centering
    \includegraphics[width=0.7\textwidth]{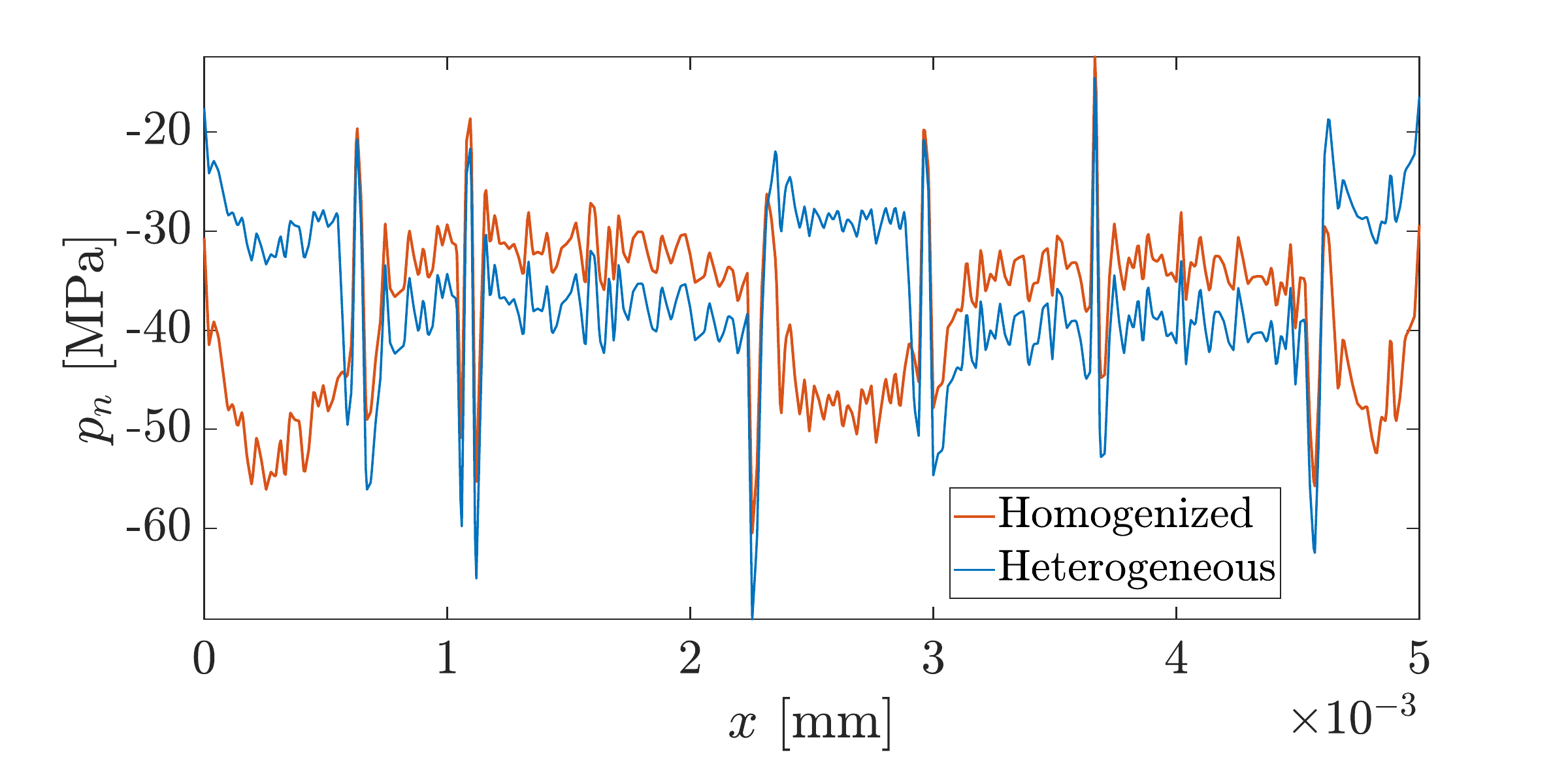}
    \caption{Contact traction $p_n$ resulting from the homogenized model having elastic modulus $E^*$, and the heterogeneous model with two different elastic modulus values for the PS and LDPE phases, for an imposed displacement of $\bar{u}=-3 h_{rms}$.}
    \label{fig:contact_stress}
\end{figure}

The heterogeneous and homogenized simulations are compared in \figref{fig:stress_domain} in terms of contour plots of the stress component $\sigma_y$ for an imposed displacement of $\bar{u}=-3h_{rms}$. The heterogeneous model clearly shows the difference between the two material phases. Moreover, the value of $\sigma_n$ in the two cases is shown in \figref{fig:contact_stress} for the same applied displacement. 

\begin{figure}[h]
    \centering
    \includegraphics[width=0.7\textwidth]{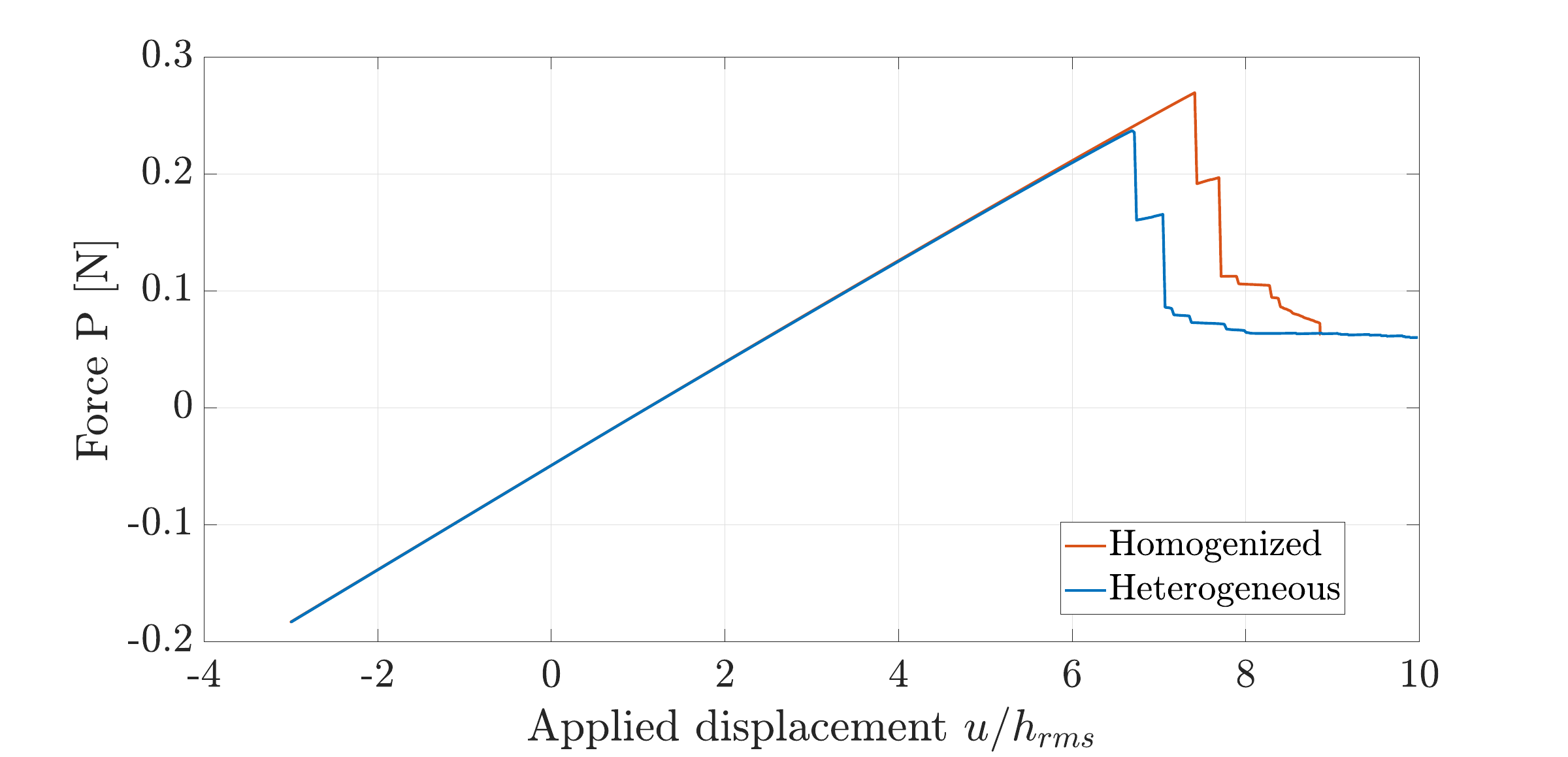}
    \caption{Comparison of the total reaction force vs. the imposed far-field displacement for the simulations with the effective elastic modulus $E^*$ (homogenized case), and with two different elastic moduli for the PS and LDPE phases (heterogeneous case).}
    \label{fig:reac_disp}
\end{figure}

Finally, the reaction forces for the two cases are shown in \figref{fig:reac_disp}. It can be noticed that, even though the two simulations refer to the same spatial distributions of the mechanical properties over the surface taken from AFM data, 
bulk homogenization performs well to predict the global average stiffness, which is coincident with the heterogeneous solution, while it gives a higher maximum force. This result shows the importance of including the material heterogeneity not only in the adhesive contact models, but also in the bulk, which is something that further motivates the use of finite element solution schemes rather than boundary element techniques. 

\FloatBarrier
\section{Conclusion}\label{sec:conclusion}

This study extended the MPJR interface finite element developed in previous works \citep{Paggi2018, Bonari2020, Bonari2022} to establish a method that incorporates experimental data from Atomic Force Microscope (AFM) into high-fidelity contact mechanics simulations, particularly for adhesive rough contacts. Integrating spatially varying adhesion properties—directly obtained from AFM measurements into the numerical model provides a more realistic and physically accurate representation of rough surface interactions at the micro- and nanoscales.

Although previous studies focused on developing and applying the MPJR method to contact problems with various surface profiles, primarily analytically or numerically generated, this work included in the contact model the actual surface topography, adhesion and dissipation maps measured by the AFM, without the need for any interpolation or other data pre-processing operation. The simulations conducted for a real PS-LDPE sample have demonstrated that the experimental data can be successfully accounted for in 2D and 3D finite element models to solve the underlying complex contact problem. The model results proved that the proposed framework successfully captures the interplay between roughness and adhesion. 

As a key research result, we have shown how to incorporate into the model the spatial variation of the adhesive properties, which are often set constant on the entire surface. By explicitly modeling these variations, we have shown in details how surface and bulk heterogeneities influence the contact predictions. In this regard, we compared solutions obtained by using heterogeneous interface properties and a homogenized bulk elasticity, and a fully heterogeneous case where also bulk elasticity is varied through the material depth. In fact, heterogeneous adhesive properties are often the result of bulk heterogeneities as well.
The results showed that the homogenized bulk approximation with a heterogeneous interface works pretty well to predict the overall stiffness of the joint, while we found that it overestimates the actual maximum force before detachment of approximately $20\%$ for the investigated PS-LDPE material combination. Hence, the fully heterogeneous case should be examined when the critical detachment load is a quantity of interest to be accurately determined, in addition to the spatial variation of the adhesive properties.



In conclusion, the proposed methodology significantly advances adhesive contact modeling, bridging the gap between experimental surface characterization from physicists and material scientists and numerical simulation from experts in computational contact mechanics.
Forthcoming studies would encompass the corresponding formulation to geometrically nonlinear and friction effects. Beyond its immediate application to adhesive rough contact problems, this methodology offers a versatile computational framework for studying a broad range of interface phenomena in tribology, material science, and nanotechnology. The MPJR framework is promising for the study of phenomena where surface roughness plays a key role, as in wear problems, designing surface topography, fracture induced by repeated application of contact loads, tire-asphalt interaction, nanoscale tribological tests based on AFM data, and multi-field tribological problems.

\FloatBarrier
\bibliographystyle{agsm}
\bibliography{bibliography}

\end{document}